\documentclass[acmsmall,screen,nonacm]{acmart}
\AtBeginDocument{%
  }

\usepackage{multirow}
\usepackage[ruled,vlined]{algorithm2e}
\usepackage[table]{xcolor}
\usepackage{fancyvrb}
\usepackage[most]{tcolorbox}
\usepackage{enumitem} % for compact itemize

\newlist{flatitemize}{itemize}{1}
\setlist[flatitemize]{nosep, leftmargin=*}

\usepackage{comment}
\usepackage{amsmath}
\usepackage{wrapfig}
\usepackage{subcaption}
\usepackage{setspace}
\usepackage{listings}
\usepackage{graphicx}
\usepackage{graphics}
\usepackage{amsmath}
\usepackage{pgfgantt}
\usepackage{pgfplots}
\usepackage{tikz}
\usepackage{amsfonts}
\usepackage{algorithmic}
\usepackage{multirow}
\usepackage{ulem}
\usepackage{color}
\usepackage{array}
\usepackage{enumitem}
\usepackage{hyperref}
\usepackage{booktabs}      % clean lines
\usepackage{longtable}     % multi-page tables
\usepackage{array}         % for p{width} column types
\usepackage{caption}  
\usepackage{adjustbox}
\usepackage{array, makecell}

\usepackage{listings}
\usepackage{xcolor}
\def\pjn{\mbox{\textsc{NSync}}}
\def\pjnagent{\mbox{\textsc{NSync-Agent}}}

\newcommand{\myauthornote}[3]{{\color{#2} \textit{\{{\sc #1}: #3\}}}}
\newcommand{\ang}[1]{{\myauthornote{Ang}{green}{#1}}}

\newcommand{\rev}[1]{{\color{black}{#1}}}

\microtypecontext{protrusion=footnote:disabled}

\renewcommand\footnotetextcopyrightpermission[1]{}
\begin{document}

%%
%% The "title" command has an optional parameter,
%% allowing the author to define a "short title" to be used in page headers.
% \title{NSync: Automated Infrastructure-as-Code Reconciliation with AI Agents}
\title{Automated Cloud Infrastructure-as-Code Reconciliation with AI Agents}

\author{Zhenning Yang}
\email{znyang@umich.edu}
\affiliation{%
  \institution{University of Michigan}
  \country{USA}}

\author{Hui Guan}
\email{huiguan@amazon.com}
\affiliation{%
  \institution{Amazon Web Services}
  \country{USA}}

\author{Victor Nicolet}
\email{victornl@amazon.com}
\affiliation{%
  \institution{Amazon Web Services}
  \country{USA}}

\author{Brandon Paulsen}
\email{bpaulse@amazon.com}
\affiliation{%
  \institution{Amazon Web Services}
  \country{USA}}

\author{Joey Dodds}
\email{jldodds@amazon.com}
\affiliation{%
  \institution{Amazon Web Services}
  \country{USA}}

\author{Daniel Kroening}
\email{dkr@amazon.co.uk}
\affiliation{%
  \institution{Amazon Web Services}
  \country{USA}}

\author{Ang Chen}
\email{chenang@umich.edu}
\affiliation{%
  \institution{University of Michigan}
  \country{USA}}

%%
%% By default, the full list of authors will be used in the page
%% headers. Often, this list is too long, and will overlap
%% other information printed in the page headers. This command allows
%% the author to define a more concise list
%% of authors' names for this purpose.
\renewcommand{\shortauthors}{Yang et al.}

%%
%% The abstract is a short summary of the work to be presented in the
%% article.
\begin{abstract}

Cloud infrastructure is managed through a mix of interfaces---traditionally, cloud consoles, command-line interfaces (CLI), and SDKs are the tools of choice. Recently, Infrastructure-as-Code/IaC frameworks (e.g., Terraform) have quickly gained popularity. 
Unlike conventional tools, IaC~frameworks encode the infrastructure in a ``source-of-truth'' configuration. 
They are capable of automatically carrying out modifications to the cloud---deploying, updating, or destroying resources---to bring the actual infrastructure into alignment with the IaC configuration.
When IaC frameworks are used together with consoles, CLI, or SDKs, IaC is unaware of changes through these non-IaC interfaces, and the IaC configuration no longer captures the intended state.  This is called \textit{infrastructure drift}. IaC frameworks will revert non-IaC changes  based on the outdated IaC configuration, leading to misconfigurations or failures.

We propose \pjn{}, an automated system for \textit{IaC reconciliation}, which aims to propagate out-of-band changes back to the IaC program in the form of an update. Our key insight is that infrastructure changes via IaC, consoles, CLI, or SDK eventually all occur via cloud API invocations---the lowest layer for cloud management operations. 
Hence, \pjn{} gleans insights from API traces to detect drift (i.e., non-IaC changes) and reconcile it (i.e., update the IaC configuration to capture the changes). 
This is a challenging task---identifying the intended change from low-level, noisy API traces is not easy; moreover, because of the criticality of cloud infrastructure, \pjn{} cannot directly test the synthesized updates in a live environment. 
\pjn{} addresses these challenges using an agentic design. It infers high-level infrastructure change intent from cloud API sequences with the help of LLMs, and synthesizes targeted IaC updates using domain-specific context management with customized agent tooling; 
it further maintains an evolving knowledge base of past successful reconciliation runs, reusing prior insights to achieve higher accuracy on future tasks. 
In addition to system design, we contribute a novel evaluation pipeline for \rev{injecting drift to cloud infrastructure} and assessing reconciliation attempts, by sourcing from authoritative cloud operation examples and transplating them into an IaC-centric framework. 
Experiments across five real-world Terraform projects and 372 drift scenarios show that \pjn{} outperforms the baseline both in terms of accuracy (from 0.71 to 0.97 pass@3) and token efficiency (1.47$\times$ improvement).

\end{abstract}

%%
%% The code below is generated by the tool at http://dl.acm.org/ccs.cfm.
%% Please copy and paste the code instead of the example below.
%%
% \begin{CCSXML}
% <ccs2012>
%  <concept>
%   <concept_id>00000000.0000000.0000000</concept_id>
%   <concept_desc>Do Not Use This Code, Generate the Correct Terms for Your Paper</concept_desc>
%   <concept_significance>500</concept_significance>
%  </concept>
%  <concept>
%   <concept_id>00000000.00000000.00000000</concept_id>
%   <concept_desc>Do Not Use This Code, Generate the Correct Terms for Your Paper</concept_desc>
%   <concept_significance>300</concept_significance>
%  </concept>
%  <concept>
%   <concept_id>00000000.00000000.00000000</concept_id>
%   <concept_desc>Do Not Use This Code, Generate the Correct Terms for Your Paper</concept_desc>
%   <concept_significance>100</concept_significance>
%  </concept>
%  <concept>
%   <concept_id>00000000.00000000.00000000</concept_id>
%   <concept_desc>Do Not Use This Code, Generate the Correct Terms for Your Paper</concept_desc>
%   <concept_significance>100</concept_significance>
%  </concept>
% </ccs2012>
% \end{CCSXML}

% \ccsdesc[500]{Do Not Use This Code~Generate the Correct Terms for Your Paper}
% \ccsdesc[300]{Do Not Use This Code~Generate the Correct Terms for Your Paper}
% \ccsdesc{Do Not Use This Code~Generate the Correct Terms for Your Paper}
% \ccsdesc[100]{Do Not Use This Code~Generate the Correct Terms for Your Paper}

%%
%% Keywords. The author(s) should pick words that accurately describe
%% the work being presented. Separate the keywords with commas.
% %% A "teaser" image appears between the author and affiliation
% %% information and the body of the document, and typically spans the
% %% page.
% \begin{teaserfigure}
%   \includegraphics[width=\textwidth]{sampleteaser}
%   \caption{Seattle Mariners at Spring Training, 2010.}
%   \Description{Enjoying the baseball game from the third-base
%   seats. Ichiro Suzuki preparing to bat.}
%   \label{fig:teaser}
% \end{teaserfigure}

% \received{20 February 2007}
% \received[revised]{12 March 2009}
% \received[accepted]{5 June 2009}

%%
%% This command processes the author and affiliation and title
%% information and builds the first part of the formatted document.
\maketitle

% each paper should have no more than 18 pages for all text and figures, plus 4 pages for references

\section{Introduction}
% Modern cloud infrastructures are managed through a mix of declarative Infrastructure-as-Code (IaC) programs and imperative interfaces such as consoles and SDKs.
Cloud management---creating, maintaining, and updating cloud resources---is a crucial task. Management tasks take place through a mix of approaches. 
Infrastructure-as-Code (IaC) frameworks are gaining popularity: Terraform~\cite{terraform} leads the market, and similar tools like Pulumi~\cite{pulumi} and OpenTofu~\cite{opentofu} are quickly rising. 
These frameworks provide declarative high level specifications of the infrastructure in readable and reusable IaC configurations. This reliable ``source of truth''  offers   
DevOps teams have a centralized view of resources, ensures reproducibility, and enables code maintenance through version control. In theory IaC frameworks should complement traditional imperative management methods such as cloud consoles, command-line interfaces (CLI), and SDKs; these imperative approaches perform direct mutative actions on infrastructure and enable quick scripting and issue remediation, but do not offer the benefits of a declarative interface \cite{cloudagent_vision}. DevOps teams would like the benefits of both: the abstract declarative view of IaC with the easily programmable, mutative power of imperative modifications.

% This multi-modal management leads to infrastructure drift—a divergence between the live cloud state and the IaC configuration—which can cause misconfigurations, deployment failures, and compliance violations.
Unfortunately, mixing IaC with imperative interfaces introduces a critical problem: \textit{infrastructure drift}, where the live infrastructure state diverges from the IaC configuration. 
Drift arises when infrastructure provisioned by an IaC framework is modified via other interfaces. An operator does this because imperative changes are easier than modifying declarative descriptions. For example, it's easier to call a single API that adds a tag to an instance than it is to find the module that describes an instance and make the modification in the appropriate place (\S\ref{subsec:drift}). 
Since IaC frameworks regard their configurations as the canonical source of truth, they will overwrite out-of-band modifications in the next IaC update. 
This can lead to serious consequences such as downtime, compliance issues, and financial penalties~\cite{sentinelone_iac}. 
% . Changes made through non-IaC interfaces (e.g., quick fixes in response to an outage) would be inadvertently undone, leading to  
% downtime, compliance issues, and financial penalties~\cite{sentinelone_iac}. 
% Rather than reverting these drifts by reapplying outdated IaC programs, we propose an alternative: IaC reconciliation—updating the IaC codebase to reflect legitimate out-of-band changes.
We propose to provide the best of both declarative and imperative worlds by
automatically incorporating out-of-band changes into the IaC configuration to consolidate the central view. We call this the \textit{IaC reconciliation} problem. 

%It inverts the traditional direction of declarative infrastructure management: instead of pushing code to update the cloud, we \textit{pull} cloud changes back into the code. This reconciliatory model embraces the multi-modal nature of modern cloud operations while preserving the core benefits of IaC -- maintainability, auditability, and reproducibility.
% This would automate a task that is today painful and error-prone, 

% Current IaC reconciliation relies on manual intervention---DevOps engineers need to discover and resolve drift manually. 
\rev{Currently, addressing the IaC reconciliation problem relies on either manual intervention or commercial tools. With manual intervention, DevOps engineers need to inspect the live infrastructure, identify discrepancies, and retrofit updates into existing IaC configurations~\cite{terraform_drift}.} 
This manual process is challenging, slow, and error-prone. While IaC frameworks can detect changes to resources they already manage (e.g., a non-IaC tool modifying an IaC-managed VM attribute), they cannot automatically discover new resources (e.g., a new subnet created via non-IaC tools). This leaves DevOps engineers to identify these unmanaged resources themselves: this is not scalable and may overlook subtle yet critical changes, leave dangling resources, and can eventually cause failed deployments and misconfigurations. 
Even if DevOps engineers successfully identify all changes, they must then translate the changes into valid IaC updates. This not only requires understanding of both the structure and logic of the existing codebase, but also mastering the correspondence between deployed cloud resources and their declarative IaC representations.

\rev{Several commercial tools have emerged to mitigate the problem. Tools like Firefly~\cite{firefly2025} and Env0~\cite{env0_cloudcompass} attempt to maintain a resource inventory by frequently scanning the entire cloud infrastructure. 
While this approach can, in theory, discover unmanaged resources, keeping the inventory accurate and up to date demands constant API calls across all regions and resource types—incurring significant operational overhead, risking API rate-limit violations, and introducing latency in drift detection.
Other tools like Spacelift~\cite{spacelift_drift} focus on managing drifts within existing IaC-managed infrastructure but cannot handle resources created outside of IaC. Moreover, these commercial solutions are not free, introducing significant operational costs to many organizations.}

We propose \pjn{}, the first automated system for IaC reconciliation. 
The key insight behind our approach is to \textit{operate on the cloud APIs} that both IaC and imperative tools rely on. 
RESTful APIs are the lowest-level abstraction for cloud management, and all higher-level modalities (IaC, CLI, SDK, consoles) invoke these APIs to perform tasks~\cite{cloudagent_vision}. 
Furthermore, cloud providers offer API monitoring services, such as AWS CloudTrail~\cite{awscloudtrail} and Azure Activity Logs~\cite{azureactivitylog}. \pjn{} uses this central vantage point to capture complete traces of all infrastructure changes, IaC or imperative. 
It jointly analyzes API traces and the original IaC configuration in order to generate code patches that reconcile the drift, while preserving the IaC configuration's structure and abstractions. 

\looseness=-1
\pjn{} casts IaC reconciliation as a \textit{program repair} task, but the domain of cloud management presents novel challenges. In traditional program repair, typically a specification or a set of test cases are given as explicit instructions for the repair.
However, in our context, the repair instructions are implicitly captured in the API traces. The repair task  needs to first perform \textit{intent identification}, summarizing long and noisy cloud API traces into a concise description of the change. 
After identifying the intent, the \textit{patch generation} step is another departure from conventional program repair. Due to the criticality of cloud operations, \pjn{} cannot push a synthesized patch into a live deployment to test correctness. 
Rather, it has to perform sophisticated static evaluations, using a set of IaC-native tools to assess the match. 
This evaluation requires nuanced reasoning across both APIs and IaC, which is difficult to capture symbolically. 
To this end, we adopt an agentic approach built on large language models (LLMs), enabling the system to interpret opaque cloud state and reason about IaC structure and style.
% We adopt an agentic approach built on large language models (LLMs). 
% A learning-based approach is needed because the cloud state is opaque yet the repair process requires a nuanced understanding of both APIs and IaC; moreover, reasoning about the IaC codebase's structure and style cannot be easily performed symbolically. 
%An agentic approach that uses a combination of neural models and symbolic analysis, 
% This is not only because the cloud is an opaque entity, but also because this repair process requires nuanced understandings at both API and IaC levels; this makes a purely symbolic approach or simply prompting an LLM ineffective. 
% Further, an agent can continuously learns from experience by noting on successful reconciliation runs and drawing from these insights for future tasks. 
Further, the agent can continuously learn from experience by retaining knowledge from successful reconciliation runs and applying these insights to future tasks.

Besides system design, another contribution we make in this paper is a novel evaluation framework for testing IaC reconciliation. We source from realistic changes that are typical for DevOps tasks, but transform these changes using an IaC-based drift injection method to obtain ground truths. We have applied this methodology to generate 372 drift scenarios across five complex, real-world Terraform projects, ranging from tens to over a thousand resources.

%Each drift scenario consists of a trace of API call events that cause modifications to the infrastructure provisioned by these projects. 
%All experiments are conducted against real deployed cloud infrastructure. 
We have built \pjn{} in an agentic framework and will release it in open source. Our evaluation shows that \pjn{} can successfully reconcile the IaC codebase,  achieving 0.95 pass@3 accuracy, outperforming the \rev{off-the-shelf Claude agent} (0.71) while being 1.72$\times$ more token-efficient. Continuous learning further improves the accuracy to 0.97 pass@3, while remaining 1.47$\times$ more efficient than baseline solutions. Under the stricter pass@1 metric, \pjn{} achieves an average accuracy of 0.80, outperforming the baseline (0.49).
To summarize, this paper makes the following key contributions:
\begin{itemize}[leftmargin=*, noitemsep, nolistsep]
    \item We present a new task called \textit{IaC reconciliation}, and devise a novel solution that gleans insights from cloud API traces and casts this task as a program repair problem.

    \item We develop \pjn{}, an agentic system that reconciles out-of-band infrastructure drift from API traces, substantially outperforming the baseline. We will release our system in open source. 
    
    \item We propose an evaluation pipeline for generating realistic, assessable drift scenarios, and curate the first IaC reconciliation dataset with 372 validated cases across five diverse Terraform projects.
    
    \item We perform a detailed empirical study to assess the effectiveness and robustness of our system across diverse drift scenarios.
\end{itemize}

\noindent In the rest of this paper, we describe these contributions in detail. 
% We evaluate \pjn{} on 371 drift scenarios across five real-world Terraform projects. Each drift scenario consists of a trace of API call events that cause distinct out-of-band modifications to the infrastructure provisioned by the Terraform projects. 
% \pjn{} successfully patches the IaC codebase in XX\% of the cases, outperforming baseline LLM agents by XX–XX\%. Its intent identification algorithm improves the accuracy by XX\%. Its domain-specific context management tools reduces input size by an average of XXX\%, resulting in XXX\% faster reconciliation time and improving reconciliation success rate by XX–XX\%. 
% These results demonstrates the importance of intent identification and context management when designing AI agent to address IaC reconciliation problem. 

% To summarize, this paper makes the following key contributions: 
% \begin{itemize}[nolistsep, noitemsep]
%     \item We formulate a novel task called IaC reconciliation; and we present a solution that updates IaC configurations based on cloud API traces.
%     \item  We develop \pjn{}, an AI system that ingests API traces to synthesize IaC patches to resolve the infrastructure drift caused by out-of-band modifications to cloud infrastructure. \pjn{} features a neuro-symbolic approach for intent identification and an agentic approach with careful context engineering for patch generation.    
%     \item We release a comprehensive benchmark of Terraform projects with induced drifts and demonstrate that \pjn{} outperforms baselines in reconciliation accuracy and efficiency.
% \end{itemize}

\section{Motivation}

%In this section, we present more background on IaC, especially the leading solution Terraform, and motivate the IaC reconciliation problem and our design of \pjn{}. 

IaC frameworks operate on a declarative view of the resources in IaC configurations. These configurations which are compiled to cloud APIs that eventually modify the infrastructure. Our work focuses on Terraform, which is the dominant IaC solution with 62\% market share as of 2025~\cite{firefly2025}. In this section, we explain how this IaC framework works, describe why engineers use a mix of interfaces, how drifts happen, and motivate the IaC reconciliation problem. Finally, we describe the challenges and solutions in automated IaC reconciliation in our \pjn{} design. 
%which stem from the difference between noisy real infrastructure information and high-level IaC representations.

\subsection{Background: IaC and Terraform}  
\label{sec:terraform}

%Developers use infrastructure-as-code (IaC) tools to manage cloud infrastructure through a declarative approach. This allows them to apply good software engineering practices such as testing, modularizing and versioning to their infrastructure management.
%Infrastructure-as-Code (IaC) enables developers to manage cloud infrastructure through declarative configuration files, applying principles of software engineering, such as testing, modularity, and versioning, to infrastructure provisioning. 

Cloud infrastructure has three forms of representation.  (1) In Terraform, DevOps engineers write \textit{configuration files} (\texttt{.tf}), which is the \textit{declarative state} that defines the desired infrastructure. Terraform codebases are often organized in the same way as large code projects, with subdirectories and multiple configuration files---e.g., each subdirectory defining a set of related resources. 
(2)~When Terraform compiles and deploys an IaC program to the cloud, it also produces a \textit{local state} file (\texttt{.tfstate}), which records the IaC-managed resources (e.g., compute instances) and their current attributes as found in the cloud (e.g., machine image, ID). This is the \textit{runtime state from the IaC point of view}, derived from the declarative state. 
(3) Finally, the cloud does not have a central view of the infrastructure state, but exposes resource-specific APIs to retrieve the \textit{remote state} per resource. 
%Finally, the cloud maintains a \textit{remote state}—the runtime view of infrastructure—possibly modified via various interfaces beyond IaC. Unlike the centralized view of IaC, remote state is usually \textit{segmented per resource} and accessed through resource-specific APIs (e.g., \texttt{GetResource}), which do not have a consist query interface.
% (3) Finally, the cloud itself also contains state about the infrastructure, called \textit{remote state}, which is the \textit{overall runtime state} that can be modified through various management interfaces, IaC or otherwise. 
%Regardless of the management interface, all cloud operations eventually take place through RESTful API calls, the lowest execution layer. 
%However, this overall cloud state is not represented explicitly, making it challenging to reconcile drift between the intended configuration and actual remote state.  
%\pjn{} addresses the challenge by monitoring the API traces that operate on the remote state, and then updating the configuration files and the local state to match with the remote state. 

\begin{figure}[t]
    \centering
    \includegraphics[width=0.99\linewidth]{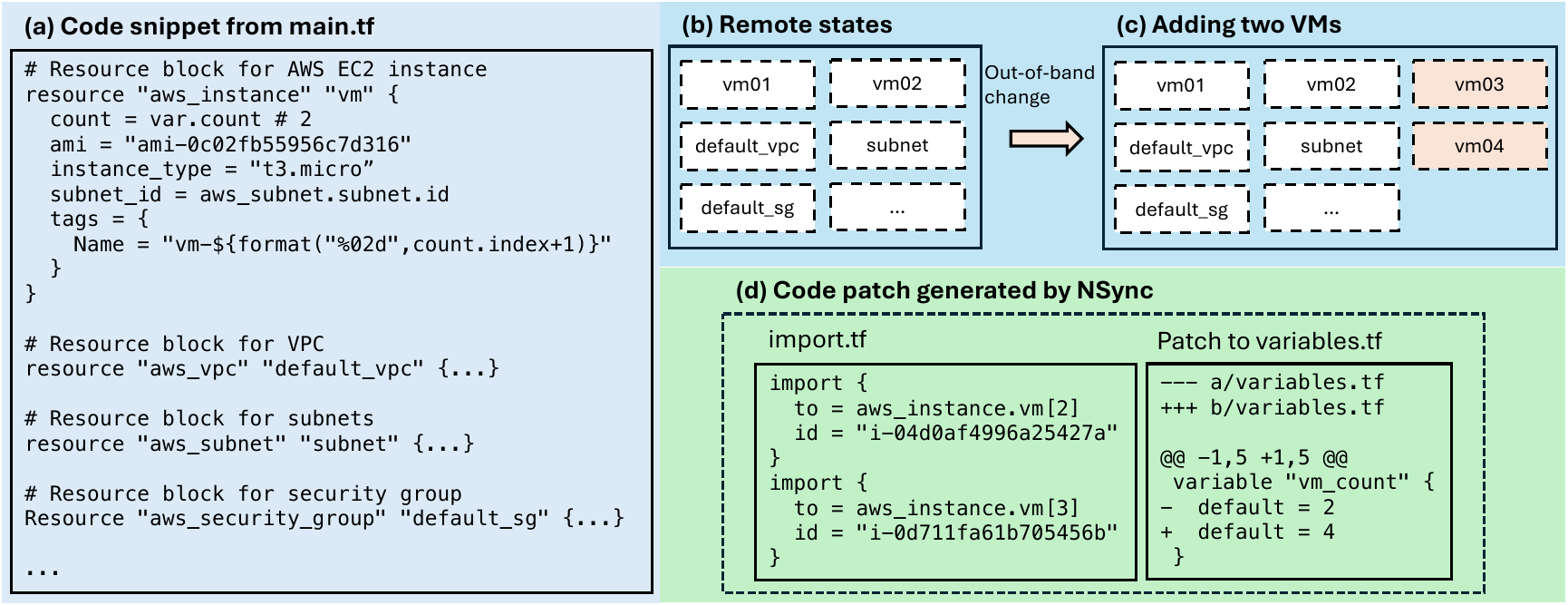}
    % \vspace{-2mm}
    \caption{A concrete IaC example written in Terraform. (a) Code snippets of a Terraform configuration file; (b)~Cloud infrastructure provisioned by the configuration files; (c) Cloud infrastructure after out-of-band change that adds two VMs; (d) Code patch generated by \pjn{}}
    \label{fig:terraform}
\end{figure}

A typical cloud operation workflow is as follows. 
First, DevOps engineers invoke \texttt{terraform plan} on a Terrraform workspace.  This will generate a set of actions that, if performed, will modify the cloud infrastructure to the state encoded in the IaC program. 
The \texttt{terraform apply} command executes the actions based on the plan, modifying the cloud's remote state.
% Note that the goal is always to modify \textit{remote state}, but Terraform only has a full view of \textit{local state} if other management interfaces are used in addition to Terraform.\victor{What does this last sentence mean?}
While IaC as ``source-of-truth'' is one of its most valued properties, it assumes that the infrastructure is entirely managed by IaC. 
Consider the example in Figure~\ref{fig:terraform}(a), which gives a Terraform configuration snippet (\texttt{main.tf})---the declarative state. This configuration provisions a number of VM instances specified by \texttt{var.count}, along with other networking components such as a Virtual Private Cloud (VPC), subnets, and a security group (SG) for the VM instances.  Figure~\ref{fig:terraform}(b) illustrates the cloud's remote state after deployment. 
% \victor{I think the rest of this paragraph is already part of Challenge \#2, we can remove, and move any part that is not said there and in challenge \#1). This subsection should focus only on explaining terraform and iac}
% In practice, unlike IaC, no central ``state file'' exists in the cloud that encodes remote state, i.e., all infrastructure resources, as the source-of-truth; rather, each resource may correspond to one or more JSON encodings, resulting in a fragmented view. 
Unlike IaC, the cloud lacks a central state file; each resource exposes its own state, yielding a decentralized and fragmented view.
In order to construct the remote state, one has to use state retrieval APIs at the resource level---i.e., one has to know that a certain VM exists before calling state retrieval APIs on that VM.

% Besides, cloud state encodings are also low-level, and markedly different from the IaC declarative definitions (Figure~\ref{fig:terraform}(a)). 
% For all of these reasons, parsing the cloud's remote state to identify the corresponding IaC resources is a daunting task. 

\subsection{Problem: Infrastructure Drift and IaC Reconciliation} 
\label{subsec:drift}

% \looseness=-1
\rev{Operations are easier if DevOps engineers only used IaC frameworks to manage their infrastructure. However, in real-world operations, they often make \textit{valid} changes outside the IaC workflow via consoles, CLIs, or SDKs. }
% After all, the latter three have been the de-facto management interfaces until the recent rise of IaC. 
% Furthermore, imperative interfaces like CLIs/SDKs present more familiar language environments, and consoles enable more direct actions via GUI automation; across the DevOps team, different engineers may also have personal preferences whether to use IaC or other tools~\cite{cloudagent_vision}. In a large company,  
% DevOps teams typically divide their tasks across subteams. Each team focuses on a different class of operational tasks and could use different tooling. 
% Hence, the team who manages the IaC frameworks and the team who changes the infrastructure (e.g., for performance scaling) may not be in sync. 
% The cloud's remote state, therefore, can be modified through all interfaces. 
% \looseness=-1
Our interviews with a major cloud provider reveal several representative situations where engineers bypass IaC workflows, which can introduce drift. 
\textit{(1) Incident response:} when troubleshooting production issues, engineers need to apply fixes immediately. To act quickly, engineers bypass IaC and use faster API-driven tools such as AWS Systems Manager Automation~\cite{awsautomation} or Azure Sentinel Playbooks~\cite{microsoftsentinel}. Drift arises when these emergency changes are never synchronized back into the IaC codebase.
\textit{(2) Performance tuning:} engineers frequently experiment with configuration parameters (e.g., scaling factors). While such modifications can be performed with IaC, doing so requires writing and deploying code for every trial, which is far more cumbersome than clicking a button in a dashboard or running a single CLI command. To iterate quickly, they bypass IaC and modify resources interactively. Drift arises when these exploratory updates are not synchronized back into IaC once an optimal configuration is chosen.
\textit{(3) Debugging and observability:} during live debugging, engineers often enable additional resources such as logging, tracing, or temporary dashboards. IaC provides little support for live diagnostics, so these are typically set up directly through consoles or APIs. Similarly, drift arises when these changes are not synchronized in the IaC codebase.

% \victor{I think the next paragraph needs a rewrite. Joey added some overleaf comments, to paraphrase (i) isn't really IaC's fault as phrased, (ii) doesn't really result in drift. I had the same thoughts.}
% Our interviews with a major cloud provider reveals several representative scenarios that can result in drift: \textit{(i)} When troubleshooting incidents in a production environment, 
% time is of the essence. Engineers prefer immediate access 
% to remediate the problem, but IaC subjects the code to a full cycle of review, testing, and CI/CD pipeline execution, which would delay critical fixes. 
% To ensure a timely response, engineers opt to use non-IaC tooling---e.g., some popular tools are AWS Systems Manager Automation~\cite{awsautomation} and Azure Sentinel Playbooks~\cite{microsoftsentinel}, which wrap around cloud APIs and can execute operational workflows more efficiently.
% \textit{(ii)} Interactive visualization and log analysis are also instrumental for live debugging, but these functions are not supported by IaC, which is primarily for resource creation. 
% % 
% \textit{(iii)} During performance tuning, engineers typically test different configurations (e.g., scaling factor) via CLI before converging on an optimal parameter. This requires spinning up, mutating, and tearing down resources repeatedly for experimentation.
% 
In all these examples, infrastructure drift could occur. 
%, which is called the IaC reconciliation problem. 
The IaC team lacks immediate visibility into when and why changes occur. Without direct operational context, they may miss relevant updates, struggle to identify which resources need to be managed or are temporary ones, and face difficulties in determining the complete scope of resource changes across different cloud services. 
Hence, infrastructure drift is fundamental to cloud operations at scale, and unfortunately, there is no fully automated solution to address the IaC reconciliation problem today. Next, we describe the three challenges in automated IaC reconciliation and our insights in \pjn{}.

%\subsection{Limitation of Existing Approaches}
\subsection{Challenge \#1: Inferring Intent for IaC Reconciliation}\label{sec:bg:infer-intent}

The first challenge is to infer what changes must be reconciled into the IaC configuration. 
%\textit{Limitation 1: Identifying out-of-band resource creation.} 
\rev{When resources are created through non-Terraform interfaces, Terraform cannot detect these changes since these resources are not part of its managed state -- they are invisible from Terraform's perspective.}
% they remain invisible to Terraform; hence, Terraform natively will not detect any drift in the infrastructure.
The current practice is that DevOps engineers need to  manually discover those resources---they must sift through SDK code commits, tickets, and verbose cloud audit logs (e.g., AWS CloudTrail); inspect multiple cloud services and regions; and attempt to understand the relationships between newly created resources and existing infrastructure. 
%The separation between teams also introduces delays in change detection, as the IaC team may not be promptly notified when infrastructure modifications occur.
 This is difficult and time-consuming, hindering operations at scale. 

The insight in \pjn{} is that modern cloud platforms provide auditing services that record \emph{all} infrastructure modifications, regardless of the interface used to perform them. 
Examples include AWS CloudTrail~\cite{awscloudtrail}, Azure Event Hub~\cite{azureactivitylog}, Google Cloud Audit~\cite{gcpauditlogs}, and Alibaba ActionTrail~\cite{alibaba-actiontrail}. 
The API call traces logged in these services offer a structured view of infrastructure modifications. Hence, these traces reflect operational behaviors, including out-of-band changes that bypass the IaC framework, providing  opportunities for automated IaC reconciliation. 
If we could interpret API traces to synthesize the corresponding IaC code updates, this will bridge the gap and translate any cloud modifications into a desired IaC update. 

\textbf{API-Driven Reconciliation.} Based on this insight, \pjn{} designs an API-driven IaC reconciliation mechanism. Recent advances in large language models (LLMs) make it feasible to infer high-level semantic intent from a collection of API calls. The challenge is to identify intent from \emph{noisy traces}. API traces are verbose and include redundant or transient operations (e.g., resources that are created and then deleted) as well as events that may be logged out of order. 
%Unlike traditional program repair where test cases explicitly specify what needs to be fixed, cloud infrastructure changes must be inferred from API traces. 
Hence, \pjn{} needs to filter out noise from API traces to identify the actual changes, and understand the intended infrastructure modifications, without explicit repair instructions (e.g., specifications, test cases).

\subsection{Challenge \#2: Generating IaC Patches from Intent}
\label{sec:bg:gen-patches}

However, even with precise knowledge of the intended infrastructure changes, patching IaC configurations to reflect those changes is challenging. 
In the cloud, there is no centralized place to query the complete infrastructure state; instead, the state is fragmented and only accessible through resource or service-specific APIs. 
Today, DevOps engineers must manually query resource states through CLI or consoles. 
These interfaces return flat, segmented descriptions populated with ephemeral runtime values, which are difficult to interpret or reuse. 
To bring these updates into the IaC codebase, engineers then need to inspect the existing configurations and manually infer not only the correct representation for each cloud resource, but also the dependencies among them.
This task is made harder by the frequent evolution of both cloud APIs and IaC schemas, which requires constant tracking and re-interpretation. As infrastructures grow larger, this workflow quickly becomes unmanageable. 
% Lifting tools such as \texttt{Terraform-cfg-gen}~\cite{hashicorp2024generating}, Terraformer~\cite{terraformer}, and aws2tf~\cite{aws2tf} attempt to reconstruct configurations by issuing these APIs calls and infer state from this fragmented information. This design makes them inherently difficult to maintain: developers must add support for each new resource or service and continuously track updates from both the cloud provider side (new APIs, changing semantics) and the IaC side (schema and language evolution), which is fundamentally unscalable. In practice, these tools produce brittle code with limited coverage~\cite{lilac}, embedding raw runtime values (e.g., identifiers, subnet IDs) instead of reusable references. The result is flat, non-modular configurations with generic names that are neither maintainable nor reusable. Crucially, they perform one-time lifting to generate new configurations, rather than producing targeted patches that respect the user’s existing IaC. For larger infrastructures, these lifting tools often degrade further, often producing syntax errors or unusable configurations~\cite{lilac}. 
These limitations motivate a more adaptive, agentic approach powered by LLMs, and we target a different goal of reconciling IaC configurations as a update patch. An LLM agent with tools can access up-to-date cloud and IaC information, generalize across resource types, and infer the structure of arbitrary user IaC codebases.

Pursuing this approach, however, requires conquering two key technical barriers. 
The first stems from the difficulty in testing an IaC patch. 
Unlike conventional code-generation settings, there is no safe way to perform live testing. 
\rev{While tools like \texttt{terraform plan} can preview potential changes, these dry-run capabilities are limited to only IaC-managed resources and cannot fully validate changes involving out-of-band resources. }
% Executing a patched IaC configuration to test correctness would directly manipulate cloud infrastructure, which can be destructive. 
At the same time, another barrier is that LLMs have limited context window, but IaC repositories can be arbitrarily large, while only a small fraction of the code is relevant to a given drift. 
These stand in contrast to conventional code-generation agents, which rely on rich environment feedback and have a scoped context---the agent can execute code, observe failures, and refine outputs using precise runtime signals as guidance~\cite{reflect_1, reflect_2, reflect_3, CodeT}. Such feedback is significantly weakened in the IaC setting.

\textbf{IaC Patch Generation without Live Testing.}  
Our design is to guide the agent by carefully managing context to isolate only the IaC fragments relevant to a drift, and developing specialized tools that help the agent extract relevant context and reason about changes safely, without manipulating live infrastructure. 
In the absence of execution feedback, the agent can rely only on our read-only IaC tools, e.g., checking IaC syntax, previewing prospective changes, for safe operations.

\subsection{Challenge \#3: Leveraging LLMs Efficiently for IaC Reconciliation}\label{sec:bg:llm-challenge}

A third challenge lies in how to efficiently leverage LLMs for reconciliation. In most domains, adapting LLMs to a specialized task relies on fine-tuning, which is  resource-intensive and poorly suited to cloud infrastructure operations---providers rapidly evolve APIs and semantics, while IaC frameworks also change frequently. Hence, fine-tuning would mean that in order to maintain accuracy, the model needs to be frequently re-trained. 
Beyond fine-tuning, recent research explores lifelong learning for agents, where systems continuously accumulate knowledge and improve through repeated interactions with their environment~\cite{voyager, trove}. These approaches have shown promise, but almost all operate in sandboxed domains, such as simulated games or controlled benchmarks, where agents can freely explore, fail, and recover at little cost. Cloud infrastructure presents a fundamentally harder setting: \rev{cloud infrastructure does not provide a safe environment for experimentation and learning.}
% every interaction is expensive, and trial-and-error directly manipulates live systems, carrying the risk of outages, destructive replacements, or security violations. 
Safe and effective lifelong learning for IaC reconciliation remains an open  challenge.  

\textbf{Learning and Reuse Across Reconciliations.}  
Our approach to enabling continual learning is based on the observation that IaC reconciliation is a repeated process.  Cloud infrastructure is long-lived, and agents may repeatedly encounter similar drift patterns as it evolves. Hence, even without free exploration, knowledge from past reconciliation attempts can be carried forward and reused safely and naturally. Specifically, the correspondence between API traces and IaC structures, or lessons from failed attempts, need not be rediscovered each time. Likewise, once an agent has learned the structure and conventions of a given codebase, such knowledge should be leveraged in future reconciliations for the same infrastructure. We design \pjn{} to incorporate prior experience, enabling safe knowledge accumulation for improving reconciliation accuracy and robustness. Our design also addresses the downside that incorporating prior knowledge enlarges context and increases inference cost, making efficiency itself a central challenge.  

% The above observations motivate a two-step approach: an \emph{intent identification} phase followed by  \emph{patch generation}, powered by LLMs. However, a na\"{i}ve approach to designing those components would lead to an inefficient solution. Existing AI agents lack specialized mechanisms for handling two key challenges that arise from the unique nature of cloud infrastructure repair: tools to validate a patch without expensive testing, and tools to accumulate specialized knowledge for better scaling. \ang{need to discuss the structure here and in sec 3.}

Together, these three challenges motivate the design of \pjn{}, an agentic system that makes IaC reconciliation safe, efficient, and continually improving over time.

\section{Design of \pjn{}}
\label{sec:design}

\pjn{} is an agentic system that uses API call traces that record cloud infrastructure events, and updates the original IaC configuration to reflect the current infrastructure state. 
\begin{wrapfigure}{r}{0.45\textwidth}
% \vspace{-2mm}
    \centering
    \includegraphics[width=\linewidth]{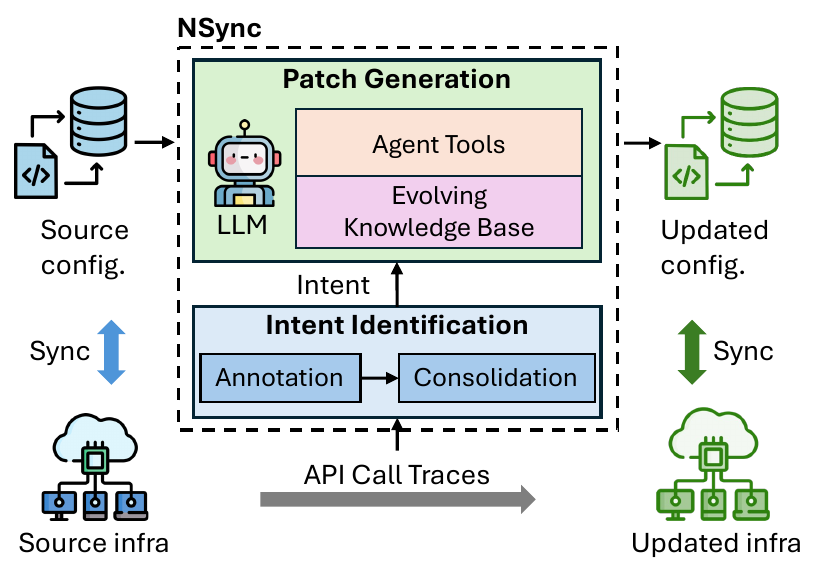}
    % \vspace{-2mm}
    \caption{Overview of \pjn{}.} 
    \vspace{-2mm}
    \label{fig:overview}
\end{wrapfigure}
Figure~\ref{fig:overview} illustrates the architecture of \pjn{}. The \textbf{intent identification} component analyzes noisy API call traces from the cloud provider to determine if there are any infrastructure changes, and analyze the nature of those changes. When infrastructure changes are detected, the second component, \textbf{patch generation}, synthesizes an update with the help of specialized IaC reconciliation 
tools, and is given access to the source configuration, the identified change intent, and a self-organized evolving knowledge base \rev{per IaC project}. It outputs updated configurations that accurately reflect the current state of the infrastructure, ensuring alignment between the IaC code and the actual deployed resources. Finally, our agent uses \textbf{continuous learning} to record past experience to achieve better performance over a longer time horizon.

\subsection{Intent Identification}
\label{sec:intent_identification}

This step aims to identify the infrastructure change intent from raw API traces. One na\"{i}ve solution is to have the LLM directly process API traces, but this presents several challenges. 
First, the traces contain substantial noise which is irrelevant to infrastructure changes, such as read-only operations, retry attempts, and user-specific account information. 
Second, API structures vary significantly across different cloud services, using inconsistent naming conventions and parameter formats, making uniform analysis difficult. 
Third, auditing services can populate API events out of order~\cite{api_outoforder_aws, api_outoforder_azure} due to asynchronous collection across services and regions and low timestamp resolution. 
This means the system should not rely on the sequential order of API calls in traces; instead, it must reason about the semantics of operations and distill the outcome of mutating actions while tolerating imprecise  arrival order. 
Our intent identification method focuses on the content and relationships of events rather than their raw sequence in the trace. 
% To address these challenges, our intent identification process works in two stages: \textit{event pruning} and \textit{event consolidation}.

Before intent identification, we preprocess the traces to extract a clean sequence of mutating events—API calls that alter infrastructure state. We discard read-only calls (e.g., \texttt{Describe*}, \texttt{List*}), deduplicate retries by keeping only the final success, and remove irrelevant fields such as timestamps or request IDs, thereby reducing trace volume while preserving all state-changing operations. The intent identification process then proceeds in two stages: \textit{annotation} and \textit{consolidation}.

The key idea is to consolidate the API events and identify the corresponding IaC-level resource blocks. Recall that imperative API calls eventually trigger changes on cloud states, which need to be encoded as IaC configuration updates. 
Inferring IaC resource-level changes (create, delete, update) from the  API calls is not easy. For instance, while the \texttt{CreateVpc} call corresponds to creating a new VPC block, a \texttt{CreateTags} call on a VPC does not create a new resource but rather updates the VPC. To capture such distinctions, we normalize API calls into a consistent schema using an LLM.

% \noindent
\looseness=-1
% \textit{\textbf{Step 1: Annotation.}} 
\subsubsection{Step 1: Annotation.}
% \zy{updated @Ang} 
Our key contribution in this step is a neurosymbolic annotation procedure that uses LLMs to produce consistent, schema-aligned labels for heterogeneous API calls. We design a fixed schema that the LLMs must adhere to; further, we encourage consistent labeling across event categories and resource types using two mechanisms: (1) the annotator maintains memory ($T$) of previously inferred resource types to ensure coherence across a trace, and (2) annotation is executed in batches with retries to guard against occasional LLM errors.

% \begin{wraptable}{r}{0.6\textwidth}
% \centering
% % \caption{Annotation schema for API calls. Fields not applicable are set to \texttt{null}.} \vspace{-2mm} 
% \caption{Annotation schema for API calls.} \vspace{-2mm} 
% \label{tab:annotation_schema}
% \resizebox{0.6\textwidth}{!}{%
% \begin{tabular}{l|l}
% \toprule
% \textbf{Field} & \textbf{Description} \\
% \midrule
% %\texttt{category} & Operation type: one of \{\texttt{create}, \texttt{delete}, 
% \texttt{category} & Operation type: \texttt{create}, \texttt{delete}, 
% \texttt{attach}, \texttt{detach}, \texttt{update} or \texttt{unknown} \\
% \texttt{type} & Primary resource type (e.g., \texttt{Vpc}, \texttt{Instance}) for create/delete/update \\
% \texttt{id} & Primary resource identifier (e.g., \texttt{vpc-123}, \texttt{i-456}), else \texttt{null} \\
% \texttt{type1} & First resource type in attach/detach relation, else \texttt{null} \\
% \texttt{id1} & First resource identifier in attach/detach relation, else \texttt{null} \\
% \texttt{type2} & Second resource type in attach/detach relation, else \texttt{null} \\
% \texttt{id2} & Second resource identifier in attach/detach relation, else \texttt{null} \\
% \bottomrule
% \end{tabular}
% }  
% \end{wraptable}

\begin{table}[t]
\centering
\caption{Annotation schema for API calls. Fields not applicable are set to \texttt{null}.}
\vspace{-2mm}
\label{tab:annotation_schema}
\resizebox{0.75\textwidth}{!}{%
\begin{tabular}{l|l}
\toprule
\textbf{Field} & \textbf{Description} \\
\midrule
%\texttt{category} & Operation type: one of \{\texttt{create}, \texttt{delete}, 
\texttt{category} & Operation type: \texttt{create}, \texttt{delete}, 
\texttt{attach}, \texttt{detach}, \texttt{update} or \texttt{unknown} \\
\texttt{type} & Primary resource type (e.g., \texttt{Vpc}, \texttt{Instance}) for create/delete/update \\
\texttt{id} & Primary resource identifier (e.g., \texttt{vpc-123}, \texttt{i-456}), else \texttt{null} \\
\texttt{type1} & First resource type in attach/detach relation, else \texttt{null} \\
\texttt{id1} & First resource identifier in attach/detach relation, else \texttt{null} \\
\texttt{type2} & Second resource type in attach/detach relation, else \texttt{null} \\
\texttt{id2} & Second resource identifier in attach/detach relation, else \texttt{null} \\
\bottomrule
\end{tabular}
}
\end{table}

\begin{wrapfigure}{R}{0.35\textwidth}
    \centering
    \resizebox{0.98\linewidth}{!}{%
        \input{algorithms/event_annotation}
    }
\end{wrapfigure}
Table~\ref{tab:annotation_schema} shows the standard schema that we define, which labels each API call into one of five canonical categories: \textit{create}, \textit{delete}, \textit{update}, \textit{attach}, and \textit{detach}. Each annotated event captures only the essential information: operation type, resource type, and unique resource identifier(s).
For example, an EC2 instance launch (\texttt{RunInstances}) is represented as \textit{create}(instance, i-1234), while a volume attachment becomes \textit{attach}(volume-1, instance-2).

Our LLM annotator reasons about the semantics of an event in context. For example, \texttt{CreateTags} may look like a new creation, but it is actually an \textit{update}, and the target of the update depends on parameters—adding a tag could modify a VM, a VPC, or another resource type. Likewise, \texttt{AuthorizeSecurityGroupIngress} appears to authorize something new, but in fact it updates the configuration of an existing security group. By contrast, \texttt{RunInstances} might seem to update the state of a VM, yet it actually provisions a new instance and should be classified as a \textit{create}. These examples show why simple regex or keyword matching would be brittle. LLMs, on the other hand, having been trained on cloud documentation including API usage examples and descriptions, understand these nuances; moreover, because cloud APIs continue to evolve, LLMs can adapt to these changes. 
To improve scalability, we annotate API calls in batches rather than individually. 
%We empirically determined that a batch size of 40 to be optimal (Sec~\ref{sec:ec_ablation}).
Algorithm~\ref{alg:annotation} takes API calls $C$ and processes them in batches $B$ of size $b$, using the schema $S$ and previously collected resource types $T$ as context. For each batch, the LLM attempts up to $r$ retries to produce annotations $\hat{B}$. The results are merged into the annotation set $A$, and the newly discovered types from $\hat{B}$ are added to $T$ for subsequent batches.

\subsubsection{Step 2: Consolidation.}
After annotation, each API call is in standardized form 
(e.g., \texttt{create(instance, i-1234)} or \texttt{attach(volume-1, instance-2)}), 
giving us a labeled trace of mutating events.
%However, this annotated trace is still noisy: it contains transient operations such as resources that were created and later deleted, redundant updates, or repeated attaches and detaches that leave no lasting effect.  
%However, reconciliation only requires reasoning about \textit{persistent drift}---the subset of operations whose effects remain visible in the infrastructure after the trace concludes. 
Reconciliation only requires reasoning about \textit{persistent drift}, i.e., the infrastructure changes that remain at the end of the trace. Relative to that goal, the trace still contains noise: there may be many transient operations, such as redundant or overwritten updates, or creation of temporary resources followed by deletion. 
% This step is not an attempt to reconstruct a full dependency graph, but rather to identify the minimal set of persistent drifts that determine how the cloud state diverges from IaC.  
Consolidation identifies the set of persistent drifts that need to be reconciled in IaC. 

Given an annotated API trace, we first organize the events by their resource identifiers. An identifier is a unique number generated by the cloud when a resource  (e.g., VM) is created, and remains consistent throughout its lifecyle. 
% ---that is, if the VM gets destroyed and another VM of the same type is recreated, it will be assigned a different identifier on runtime. Edges, on the other hand, are identified by the two nodes/resources that it connects. 
For each \textit{node identifier}, we collect all events that occurred on this node/resource---e.g., \texttt{subnet-456: [create, update]} means that a unique ID \texttt{subnet-456} was created and then updated. 
For each \textit{edge} that connects two resources, we collect all operations that have been performed on this edge, written as \texttt{$\langle$id1, id2$\rangle$}. 
For instance, \texttt{$\langle$volume-1, instance-2$\rangle$: [attach, detach, attach]} denotes three operations on this edge. 
We then apply the following reduction rules independently to each node or edge identifier:  

\begin{itemize}[leftmargin=*, noitemsep, nolistsep]
    \item \textbf{Persistent Create.} 
    If a resource is created and never deleted, we retain only the create event and drop intermediate updates. Creates take precedence as they introduce new resources, while configuration details can be recovered later.

    \item \textbf{Persistent Delete.} 
    If a resource was deleted, 
    we retain only the delete and remove all associated updates prior to the deletion. 
    Deletes, like creates, are higher-precedence persistent drifts, 
    as they eliminate resources entirely from the infrastructure.  

    \item \textbf{Persistent Update(s).} 
    If a resource was only updated (with no create or delete), we conservatively retain the update. 
    This ensures that updated attributes, even if they are not defined in the original  IaC configuration, will be explicitly captured.
    % This ensures that attributes missing from the IaC code---including defaults that \texttt{terraform plan} would silently accept---are explicitly captured, preventing hidden inconsistencies in future deployments.
    % If a resource was only updated (with no create or delete), we conservatively retain the update. 
    % This case implies the resource already existed prior to the trace and is therefore managed by IaC. 
    % While IaC tooling such as \texttt{terraform plan} can detect many of these drifts, it checks attributes explicitly defined in the IaC code. 
    % When updates involve attributes with default values that were not declared in the configuration, \texttt{terraform plan} silently updates the state file without flagging a drift. 
    % This creates hidden inconsistencies: if the same configuration is applied in a fresh environment, Terraform will provision resources with wrong default values again. 
    % By retaining such updates, our agent can add the missing attributes into the IaC code, ensuring it faithfully reflects the actual cloud state.

    \item \textbf{Persistent Attach/Detach.} 
    Balanced attach/detach pairs cancel out, while any net imbalance is retained as a conservative signal of persistent drift. 
    %We do not attempt to determine the exact final relationship state from the trace itself, 
    %since event ordering may be unreliable. 
    %Instead, our goal is to flag potential drift; the precise state of relationships can later be confirmed 
    %through safe, read-only operations invoked by the agent during reconciliation.  
\end{itemize}  
The output of this reduction is a compact set of persistent drifts, organized by resource and edge IDs, and it is deliberately conservative to avoid discarding meaningful information. 

\subsection{Patch Generation} 

% Next, \pjn{} generates a patch to fix the persistent drift. 
% Generating IaC patches with LLMs presents unique challenges compared to traditional 
% automated program repair (APR). 
% In APR, repair is guided by execution feedback: candidate patches 
% are applied, errors or test failures are observed, and the process is refined step by step. 
% However, for cloud management, directly interacting with the environment to converge on a correct solution is dangerous since agents may damage the cloud infrastructure by accident. 
% Hence, \pjn{} replaces live execution with safe, read-only operations to address the unique challenges of cloud operations. 
% Our novel contributions are the design IaC-specific agent tools that provide feedback without modifying infrastructure, allowing the agent to iteratively refine patches in a safe manner. 
%To support this workflow, we design specialized IaC-specific agent tools:  

\rev{
Next, \pjn{} generates a patch to fix the persistent drift. Generating IaC patches with LLMs presents unique challenges compared to traditional automated program repair (APR). In APR, repair correctness can be validated through test execution and runtime signals. However, for cloud management, validating patch correctness is more challenging. Direct execution of a patched IaC project is not feasible as it would modify live infrastructure. While tools like \texttt{terraform plan} can preview potential changes after a patch is applied, it is designed for previewing planned infrastructure updates from a known terraform state to a new desired state, rather than validating drift reconciliation patches where resources have been created or modified outside of terraform's knowledge. This makes it difficult to assess patch correctness before actual deployment.
}

% Its feedback can be misleading since it only considers terraform-managed resources and cannot reason about drifted resources that exist outside of terraform's management. 

\begin{itemize}[leftmargin=*, noitemsep, nolistsep]
    \item \texttt{drift\_report}: a safe, read-only operation that previews the alignment between the 
    current patch and the cloud state (inspired by  \texttt{terraform plan}; see later).
    % , allowing the 
    % agent to evaluate candidate patches without modifying infrastructure.  

    \item \texttt{self\_critique}: inspired by reflection and self-refinement methods~\cite{reflect_2, reflect_3}, 
    this tool lists all edits made so far and requires the agent to reason about their correctness and necessity. 
    By critiquing its own patch history, the agent avoids hallucinated changes, scope creep, and untracked 
    divergences from the IaC codebase.  
\end{itemize}  
The patch generation workflow in \pjn{} is an iterative patch–evaluate–refine loop. The agent first generates a candidate patch, then evaluates it using feedback tools, and finally critiques its own edits before refining. This mirrors the trial-and-error workflow of program repair agents, but here it is realized entirely through safe, read-only operations tailored to IaC.

\subsubsection{Drift Report}\label{sec:drift_report}
%Within this loop, one central feedback channel is \texttt{drift\_report}. 
% Existing \texttt{terraform plan}-based drift reporting tools~\cite{terraform, cdk_drift, pulumi_drift} are ill-suited for agents. 
% \rev{\texttt{Terraform plan} reports drifts but is ill-suited for agents. Its output is verbose and often misleading.}
\rev{\texttt{Terraform plan} is not suitable for patching because it assumes the out-dated IaC configuration is the ``source of truth'' and generates update plans that propose reverting cloud-side changes to match stale configurations}
% For instance, the resulting plans enumerate every attribute difference in exhaustive detail, inflating context length and overwhelming LLMs. 
% Worse, because they assume the IaC configuration is the ``source of truth,'' they generate update plans that propose reverting cloud-side changes to match stale configurations. 
An agent reading this information effectively has to ``reverse engineer'' the changes with substantial ingenuity. 
\rev{Worse, \texttt{Terraform plan} exposes attribute-level differences for each managed resource, including runtime fields marked as ``known after apply,'' resulting in verbose outputs that inflate context length and overwhelming LLMs.}
This combination of misleading guidance and excessive verbosity effectively poisons the agent’s reasoning, pushing it toward incorrect or irrelevant patches.

\texttt{drift\_report} reframes planning into reconciliation-oriented feedback. 
Instead of emitting a full update plan, it trims outputs to only those resources that actually drifted 
and annotates them with their locations in the IaC codebase. 
\rev{This targeted feedback helps \pjn{} generate and refine patches more effectively. The agent first analyzes the condensed API trace to create an initial patch, then uses \texttt{drift\_report} in an iterative patch-evaluate-refine loop to improve the solution. Currently, \texttt{drift\_report} assumes all detected infrastructure changes should be incorporated into the IaC patch. 
However, in practice, some drifts are intentional and should be preserved, while others should be reverted. 
Distinguishing between them remains future work, as this study focuses on automatically reconciling all drifted resources under IaC management so that DevOps engineers can subsequently perform operations directly through IaC.
}

\subsubsection{Self Critique}
Another challenge in the agentic workflow is maintaining focus across multiple edits. After reasoning over the drift and applying a sequence of patches, the agent can lose track of its overall reconciliation objective, leading to hallucinated changes or oversized edits that diverge from the intended fix~\cite{liu-etal-2024-lost}. Prior work suggests that reflection can improve agent reliability~\cite{reflect_1,reflect_2,reflect_3}, but existing methods depend on execution feedback or test cases—signals unavailable in IaC reconciliation, where live execution is unsafe.

\looseness=-1
We design \texttt{self\_critique}, a safe reflection mechanism that operates without execution, which operates at a coarser granularity than \texttt{drift\_report}. 
Whereas \texttt{drift\_report} helps the agent verify whether a candidate patch has fully reconciled the infrastructure by checking if any drift remains.
\texttt{self\_critique} periodically reviews the accumulated edits and prompts the agent to reason about whether its progress still aligns with the drift intent. By surfacing reconciliation at this higher level, the tool enforces consistency and helps the agent stay focused on the reconciliation objective.

\subsection{Continual Learning} %\ang{make consistent}
\label{sec:agent_learning}

Our next design is on continual learning. 
% A na\"{i}ve reconciliation agent would treat each drift scenario in isolation and effectively ``forget everything'' once a session ends. However, 
Cloud infrastructure is long-lived and reconciliation tasks repeated arise. The lack of memory would force the agent to repeatedly rediscover mappings between API traces and IaC resources, relearn project-specific conventions, and re-synthesize fixes for error cases it has already encountered. Such redundancy not only wastes compute resources but also limits robustness, since reconciliation of large IaC projects often involves recurring drift patterns that demand consistency across runs. 
To overcome this limitation, we introduce a lightweight, domain-specific knowledge base (KB) that enables continual learning~\cite{trove, voyager} across reconciliation sessions. The KB captures information about prior reconciliations, such as effective patching strategies, frequently observed drift patterns, and project-specific semantics (e.g., naming conventions, module structures, or dependency layouts). This KB is continuously updated after a reconciliation task completes; during the next reconciliation, the agent retrieves relevant knowledge on-demand to guide patch generation, effectively grounding its reasoning in accumulated experience rather than starting from scratch.

We design the KB using the following principles. (1) Updates are \textit{selective}: only reconciliations that succeed contribute new entries, while failed sessions are discarded to avoid propagating errors.
(2) The KB is \textit{scoped per project} rather than shared globally, since much of the knowledge (e.g., naming conventions, module structure) is project-specific; this prevents cross-project contamination while allowing repeated reconciliations on the same project to benefit from accumulated insights. 
The learning process is 
\textit{agent-driven}: the agent explicitly invokes the tools \texttt{knowledge\_update} and \texttt{knowledge\_retrieval} when it judges that the current context is complete and accurate enough to warrant persistence for reuse, as opposed to always logging every intermediate action. 
Constructing a useful KB entry requires the complete and up-to-date context of a reconciliation attempt; otherwise, the agent risks recording incomplete or misleading knowledge. 
We employ a ReAct-style~\cite{yao2022react} prompting strategy that enforces an \textit{Observation}--\textit{Thought}--\textit{Action} loop on every iteration, instructing  the agent to ground its reasoning in the latest observations before deciding whether to update or retrieve knowledge.  
For example, if the agent repeatedly fails to apply a patch because an S3 bucket update requires adjusting its versioning setting, once it discovers the correct fix it can invoke \texttt{knowledge\_update} to record this resolution for future runs.  
Over time, this mechanism enables a form of experience-guided reconciliation, where the agent progressively adapts to each IaC project. 

\rev{
The KB is a light-weight, project-specific text file for our prototype. It can be easily extended with vector database~\cite{faiss} support for larger or more complex knowledge.
The \texttt{knowledge\_retrieval} tool queries the KB to access previously recorded reconciliation insights, while \texttt{knowledge\_update} writes new successful reconciliation experiences to the KB. 
This lightweight implementation ensures that knowledge persistence and retrieval remain simple yet effective, while maintaining the per-project scoping principle.
}
The agent has the ability to edit or discard outdated KB entries when previously successful strategies fail under evolving external systems, ensuring that knowledge remains accurate over time.

\section{Evaluation}
\label{sec:eval}

We conduct a comprehensive evaluation of \pjn{} on real-world Terraform projects and realistic drifts. We design our experiments to answer five key research questions:

\begin{itemize}[leftmargin=*, noitemsep, nolistsep]
\item \textbf{RQ1 (Effectiveness) + RQ2 (Efficiency)}: How effective is \pjn{} at reconciling infrastructure drifts, as measured by pass@k accuracy, compared to a baseline agent? How well does \pjn{} reduce computational overhead  while maintaining or improving reconciliation accuracy?

\item \textbf{RQ3 (Intent Identification)}: How well does API trace consolidation help identify the intent, thus improving performance, and what are the trade-offs in accuracy and efficiency? 

\item \textbf{RQ4 (Patch Generation)}: How do specialized IaC tools improve reconciliation performance, and what does their usage frequency and timing reveal about the agentic problem-solving process? 

\item \textbf{RQ5 (Continual Learning)}: 
% How robust is reconciliation under stricter pass@1 evaluation without retries, and 
\rev{How reliably does the learning algorithm improve outcomes?} 
%under arbitrary ordering of drift scenarios?

\end{itemize}

\noindent In the rest of this section, we first describe our methodology in curating realistic drifts and evaluating reconciliation effectiveness, discuss the experimental setup, and then answer these questions. 

\subsection{Evaluation Methodology: Generating and Evaluating Realistic Drifts} 
\label{sec:eval_benchmark}

\looseness=-1
Drift reconciliation is a novel task, which requires a new methodology for evaluation. Our evaluation pipeline addresses two challenges: (1) generating realistic drifts scenarios and (2) evaluating them against ground truth. This pipeline and dataset are another contribution of our paper. 

\looseness=-1
\textbf{Realistic drift scenarios.} 
To evaluate reconciliation performance, we need realistic drift scenarios that represent how DevOps engineers modify cloud infrastructure in practice. We source these scenarios primarily from AWS Systems Manager (SSM) Automation documents~\cite{aws2024runbook}, which are AWS-provided operational runbooks containing both natural language descriptions and executable API scripts for common infrastructure management tasks. For example, one such scenario is ``enabling CloudWatch monitoring on EC2 instances.'' By using these authoritative AWS documents, we ensure our injected drifts reflect real-world infrastructure changes. We further supplement these AWS-sourced scenarios with manually curated drifts specific to each IaC project. \rev{The complete list of SSM-sourced scenarios and project-specific scenarios can be found in Appendix~\ref{sec:appendix-runbook} and Appendix~\ref{sec:appendix-custom-mutation} respectively.}

% In order to evaluate reconciliation performance, we need to generate realistic drift scenarios---representative IaC changes that DevOps engineers might make to real-world projects. 
% We sample \textit{drift scenarios} from AWS Systems Manager (SSM) Automation documents~\cite{aws2024runbook}. 
% These documents are common operational tasks provided by AWS, and contain natural language descriptions of concrete actions, as well as API-calling scripts for performing these actions; for instance, an example scenario would be ``enabling CloudWatch monitoring on EC2 instances.''
% Sourcing from authoritative documents, \pjn{} ensures that injected drift closely mimic changes that DevOps engineers would make to their infrastructure. 
% \rev{Appendix~\ref{sec:appendix-runbook} lists the sampled drift scenarios from SSM. To supplement the drift scenarios, we also manually curated a few drift scenarios tailored to each IaC project. \rev{Appendix~\ref{sec:appendix-custom-mutation} lists the manually curated drift scenarios per project. }

\begin{figure}[t]
    \centering
    \includegraphics[width=0.99\linewidth]{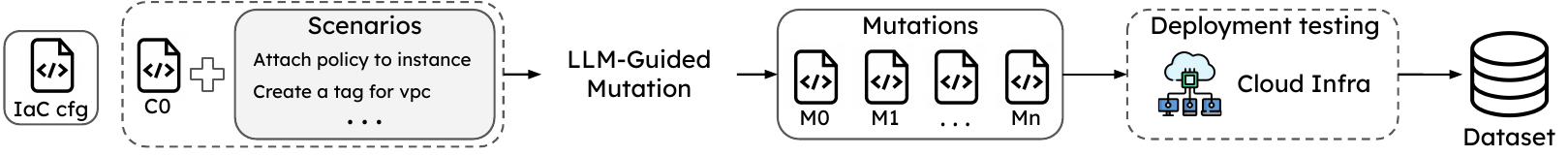}
    % \vspace{-3mm}
    \caption{Dataset generation pipeline. Mutated configurations are generated from a base configuration and natural-language scenario, validated through a deployment testing, and retained if successful. }
    \label{fig:dataset_collection}
\end{figure}

% \looseness=-1
\textbf{Generating assessable drifts.}
The next challenge arises in generating drift in a manner that allows reconciliation to be evaluated. 
One na\"{i}ve strategy is to induce drift by simply executing the API scripts of the drift scenarios in the IaC-managed infrastructure. 
However, this approach is difficult to evaluate: there is no \textit{ground truth} against which we can compare the reconciled IaC configuration, short of manually inspecting the new IaC configuration and the cloud state, which would not permit scalable evaluation.  
We design a novel dataset generation pipeline (Fig.~\ref{fig:dataset_collection}). which analyzes a drift scenario using LLMs, but generates a validated \textit{IaC configuration} capturing the mutations; it further validates the mutation by attempting to deploy it. During the generation process, we reset the infrastructure back to its original state after each mutation---matching the base configuration \texttt{C0}---manually repairing the infrastructure if necessary.
\rev{We inject drift by deploying mutated IaC configurations rather than directly executing API scripts. This approach enhances reproducibility and enables automated validation of patch correctness. First, for reproducibility, API scripts typically depend on existing resource identifiers and runtime parameters from the original infrastructure, making them difficult to reproduce since these identifiers cannot be known beforehand and must be resolved during execution. In contrast, IaC configurations declaratively specify the desired state, making drift injection reproducible across different environments. Second, for validation, having the mutated IaC configuration provides a ground truth against which we can automatically evaluate patch correctness - we can compare the reconciled patch against the known desired infrastructure state derived from the ground truth configuration.}

\textbf{Evaluation pipeline.} The inputs of the evaluation pipeline are a base workspace containing the IaC configuration and a mutated configuration. The configuration in the base workspace is deployed, generating a local Terraform state.
%When evaluating \pjn{}, we first deploy the base configuration in a ``base workspace,'' which generates a cloud deployment and a local Terraform state. 
Then, in a separate ``drift workspace'', we induce drift by applying the mutated configuration to the same underlying cloud infrastructure. 
Since the two workspaces are isolated, the base workspace is unaware of the out-of-band changes. 
We use CloudTrail to capture API events, and provide this trace together with the base workspace to \pjn{} for drift reconciliation.
\rev{To validate patch correctness, we compare the patched configuration against the ground truth infrastructure state using \texttt{terraform plan}. The ground truth is derived from the mutated configuration and represented as a local state file. 
A patch is considered correct when it satisfies two conditions: 
(1) all resources that previously existed outside of Terraform are now defined in the configuration and are ready to be imported to the local state file via ``import'' actions; and
(2) all Terraform-managed resources match their desired state—meaning the configuration and state are consistent, and any deleted resources have been removed from the configuration.
If \texttt{terraform plan} reports no changes other than the required ``import’’ actions, the patch is considered correct. Any additions, updates, or deletions indicate that the patch failed to fully reconcile the infrastructure with the ground-truth state.
}
% During dataset generation, we exclude cases where mutated infrastructures produce errors. 
In evaluation, when an agent produces the correct resource under a different name, we detect the match and insert \texttt{moved} blocks to avoid false negatives.
A \texttt{moved} block informs Terraform of differently aliases of the same resource, preventing spurious delete–create actions.

\begin{table}[t]
\caption{Benchmark statistics: configuration size, mutation space, and event counts.}
\vspace{-2mm}
\label{tab:benchmarks}
\centering
\resizebox{0.88\textwidth}{!}{%
\begin{tabular}{r|c c c|cc|cc}
\toprule
\multirow{2}{*}{\textbf{Benchmark}} 
& \multirow{2}{*}{\textbf{\#Resources}} 
& \multirow{2}{*}{\textbf{LoC}} 
& \multirow{2}{*}{\textbf{\#Scenarios}} 
& \multicolumn{2}{c|}{\textbf{\#API Events}} 
& \multicolumn{2}{c}{\textbf{\#Mutating Events}} \\
\cmidrule(lr){5-6} \cmidrule(lr){7-8}
& & & & Avg & (Min, Max) & Avg & (Min, Max) \\
\midrule
lab12~\cite{AWSAdvancedNetworking} & 47   & 289   & 96 & 183.2 & (90, 441) & 9.5 & (1, 47) \\
flask~\cite{terraform_microservice} & 74   & 799   & 94 & 280.8 & (158, 671) & 10.0 & (1, 56) \\
ssm3~\cite{terraform_automation}   & 66   & 377   & 103 & 268.0 & (149, 737) & 8.0 & (1, 48) \\
live-score~\cite{terraform_livescores} & 193  & 1582  & 61 & 612.3 & (373, 1163) & 8.5 & (1, 42) \\
mega-mesh~\cite{terraform_megamesh} & 1930 & 7087  & 18 & 3735.9 & (3692, 3920) & 7.8 & (1, 31) \\
\bottomrule
\end{tabular}}
\end{table}

% Stats for ../_nsync_workspace_baseline/aws-advanced-networking-lab12_results.csv:
% API Total - min: 90, max: 441, mean: 183.17
% Mutating Count - min: 0, max: 47, mean: 9.46

% Stats for ../_nsync_workspace_baseline/terraform-examples-ssm-automation-3_results.csv:
% API Total - min: 149, max: 737, mean: 268.02
% Mutating Count - min: 0, max: 48, mean: 8.01

% Stats for ../_nsync_workspace_baseline/flask-microservices_results.csv:
% API Total - min: 158, max: 671, mean: 280.76
% Mutating Count - min: 0, max: 56, mean: 10.03

% Stats for ../_nsync_workspace_baseline/live-scores_results.csv:
% API Total - min: 373, max: 1163, mean: 612.25
% Mutating Count - min: 0, max: 42, mean: 8.52

% Stats for ../_nsync_workspace_tools/mega-mesh_results.csv:
% API Total - min: 3692, max: 3920, mean: 3735.90
% Mutating Count - min: 0, max: 31, mean: 7.75

\textbf{Dataset contribution.} 
Using this pipeline, we have curated the first dataset of realistic IaC drift scenarios for systematic evaluation of reconciliation agents. Our five base projects are public Terraform respositories that vary widely in size and domain, from tens to thousands of resources (Table~\ref{tab:benchmarks}). 
They include \textbf{lab12}~\cite{AWSAdvancedNetworking} (multi-VPC networking), \textbf{ssm}~\cite{terraform_automation} (zero-downtime patching), \textbf{livescore}~\cite{terraform_livescores} (event-driven service), \textbf{flask}~\cite{terraform_microservice} (REST API with DynamoDB), and \textbf{megamesh}~\cite{terraform_megamesh} (multi-region VPC mesh). Our dataset contains 372 validated drift cases, generated by mutating and deploying these real-world Terraform projects.

\subsection{Experimental Setup} 

\begin{wraptable}{R}{0.35\linewidth}
% \vspace{-2mm}
\centering
\caption{Tools available to all agents.}
\vspace{-2mm}
\label{tab:tools}
\resizebox{0.99\linewidth}{!}{%
\begin{tabular}{ll}
\toprule
\textbf{Agent Tools} & \textbf{Description} \\
\midrule
\texttt{file\_read}    & Read a file \\
\texttt{file\_write}   & Overwrite a file \\
\texttt{editor}        & Fine-grained edits \\
\texttt{shell} & Run shell commands \\
\bottomrule
\end{tabular}
}
% \vspace{-3mm}
\end{wraptable}

We implemented \pjn{} in 11k lines of code in Python, and it targets Terraform-based IaC for the AWS cloud; it uses Boto3 to interface with CloudTrail for obtaining API traces.

% To the best of our knowledge, there are no existing automated systems that directly address the IaC reconciliation problem---i.e., updating declarative infrastructure code based on out-of-band procedural changes captured via API traces. 
% Infrastructure “lifting” tools such as Terraformer or aws2tf are excluded from comparison, as they generate new configuration code but do not repair existing codebases or operate over procedural traces. \ang{one potential objection is to use Terraformer to generate full lifting results, and then prompt llms to trim down to a repair. so we need a stronger reason, or simply be silent about lifting tools.}

\textbf{Baselines.} 
We evaluate three agents, all built using the Strand agent framework~\cite{strandsagents} and deployed via the AWS Bedrock service with Claude 3.7 Sonnet as the backbone LLM. 
The Baseline solution is a LLM agent equipped with pre-built file editing tools and shell access, but without domain specialization. 
The second method, \pjn{}-NL, is our system without continual learning, where each reconciliation run is treated independently. 
Our full system, \pjn{}, builds on the same framework and backbone, but with all domain-specific mechanisms detailed in Section~\ref{sec:design}---including intent identification, patch generation (e.g., \texttt{drift\_report} and \texttt{self\_critique}), as well as a project-level knowledge base (KB) that enables continual learning across runs. 
The annotation process in both variants relies on the same Claude 3.7 Sonnet model.
All three systems operate on identical inputs (mutating API traces and the IaC project directory) and share the same core editing tools shown in Table~\ref{tab:tools}.
% All methods operate on an API trace that has been pruned to retain only mutating events. 
All experiments are conducted against realistic deployments on AWS to ensure results reflect practical operating conditions. 
For each system, we perform three independent runs, with drift scenarios presented in arbitrary order to reduce any bias from experiment ordering.  

\textbf{Metrics.} 
We focus on correctness and efficiency, using the pass@$k$ metric. 
A reconciliation attempt is considered correct if the agent’s patched configuration, when evaluated with \texttt{terraform plan} against the \textit{ground truth state} and the \textit{live infrastructure} created by the mutation, produces no differences---for example, the out-of-band creation of a new EC2 instance has been captured by the patch, i.e., it includes the corresponding resource block (e.g., \texttt{aws\_instance}).  
Efficiency is measured in terms of two complementary metrics: tokens processed (computational cost), agent steps. 
Since multiple candidate patches may be generated for each scenario, we report results using the pass@$k$ metric, which measures whether at least one of the top-$k$ patches is correct while accounting for the associated efficiency cost. 
We report both pass@1 and pass@3 results to evaluate the single-attempt accuracy of the system and its robustness under limited retries.

\subsection{RQ1 {\&} 2: Effectiveness and Efficiency}
\label{sec:overall_result}

\begin{table}[t]
\centering
\caption{Accuracy and Efficiency Comparison between Baseline, \pjn{}-NL, and \pjn{}. M: Millions.}
\vspace{-2mm}
\resizebox{0.99\textwidth}{!}{%
\begin{tabular}{rc|ccc|ccc|ccc}
\toprule
\multirow{2}{*}{\textbf{Benchmark}} & \multirow{2}{*}{\textbf{\# Exp.}} &
\multicolumn{3}{c}{\textbf{Pass@3 Accuracy}} &
\multicolumn{3}{c}{\textbf{Avg Tokens (M)}} &
\multicolumn{3}{c}{\textbf{Avg Steps}} \\
\cmidrule(lr){3-5}\cmidrule(lr){6-8}\cmidrule(lr){9-11}
& & Baseline & \pjn{}-NL & \pjn{} & Baseline & \pjn{}-NL & \pjn{} & Baseline & \pjn{}-NL & \pjn{} \\
\midrule
lab12 & 96 & 0.89 & 0.98 & \textbf{0.99} & 0.45 & \textbf{0.26} & 0.36 & 21.7 & \textbf{14.5} & 16.2 \\
flask & 94 & 0.64 & 0.95 & \textbf{0.97} & 0.73 & \textbf{0.57} & 0.66 & 26.1 & \textbf{18.2} & 19.4 \\
ssm3 & 103 & 0.74 & 0.92 & \textbf{0.97} & 0.52 & \textbf{0.32} & 0.33 & 21.0 & \textbf{15.3} & 15.5 \\
live-score & 61 & 0.57 & 0.93 & \textbf{0.98} & 0.70 & \textbf{0.39} & 0.51 & 22.4 & \textbf{17.3} & 21.6 \\
mega-mesh & 18 & 0.72 & \textbf{0.94} & \textbf{0.94} & 1.05 & \textbf{0.44} & 0.50 & 18.9 & 18.1 & \textbf{15.6} \\
\midrule
\textbf{Overall} & \textbf{372} & 0.71 & 0.95 & \textbf{0.97} & 0.69 & \textbf{0.40} & 0.47 & 22.0 & \textbf{16.7} & 17.7 \\
\bottomrule
\end{tabular}}
\label{tab:detailed_benchmark_comparison}
\end{table}

% To answer RQ1 and RQ2, \pjn{} proves highly effective, significantly outperforming the baseline agent. 
% Across 372 diverse real-world projects and mutations, \pjn{} achieves 0.97 accuracy, far exceeding the baseline. 
% At the same time, it is more efficient, consuming on average 0.47M tokens per reconciliation, a 1.47$\times$ improvement in token efficiency.

% Table~\ref{tab:detailed_benchmark_comparison} summarizes the accuracy and efficiency results of \pjn{}, \pjn{}-NL and the baseline. 

% \pjn{}-NL and \pjn{} significantly outperform the baseline across all benchmarks. Overall, \pjn{}-NL improves pass@3 accuracy from 0.71 to 0.95, a relative gain of about 34\%, while also reducing token usage by more than 40\%. 
% The gains are most pronounced on challenging projects such as \textit{flask} (0.64~$\rightarrow$~0.95) and \textit{live-score} (0.57~$\rightarrow$~0.93), where baseline agents struggle to reconcile accurately. 
% \pjn{} further boosts accuracy to 0.97 by reusing knowledge from prior runs, albeit with a small increase in tokens and runtime compared to \pjn{}-NL. 
% Both systems remain far more efficient than the baseline, requiring fewer reasoning cycles and less computation overall. 
% These results demonstrate that domain specialization yields substantial accuracy and efficiency benefits, while learning mechanisms provide an additional layer of robustness.  

We start by answering RQ1 and RQ2, and Table~\ref{tab:detailed_benchmark_comparison} summarizes the detailed accuracy and efficiency results of \pjn{}, \pjn{}-NL, and the baseline.  \pjn{} proves highly effective, significantly outperforming the baseline agent. 
Across 372 diverse real-world projects and mutations, \pjn{} achieves 0.97 accuracy, far exceeding the baseline. 
At the same time, it is more efficient, consuming on average 0.47M tokens per reconciliation, a 1.47$\times$ improvement in token efficiency.
Furthermore, both \pjn{}-NL and \pjn{} significantly outperform the baseline across all benchmarks, requiring fewer reasoning cycles and less computation overall. 
On simpler projects (e.g., \texttt{lab12}), which has a flat structure without modules, all methods perform well. The performance gap widens on projects with custom or prebuilt modules, where the code structure is more complex. 
And the performance of \pjn{}-NL and \pjn{} scale more gracefully to project complexity. 

A closer inspection shows that the baseline consumes more tokens and fails more often, because it relies heavily on raw \texttt{terraform plan} outputs. Its context becomes cluttered with verbose and often irrelevant information produced by this Terraform-native tool. 
The context window overflows, triggering summarization that discards useful details. 
This forces the agent into repeated reasoning and revisiting Terraform files, further increasing token usage. 
Even worse, the baseline hallucinates: due to the bloated context dominated by \texttt{plan} noise, it incorrectly reasons that applying the current configuration will fix the drift (i.e., undoing the out-of-band change), rather than updating the configuration to reflect the changes. 
In contrast, \pjn{}'s specialized \texttt{drift\_report} tool feeds only the relevant drifted resources into the model’s context. 
This focused signal not only improves efficiency but also yields higher accuracy by avoiding context dilution.

\begin{wraptable}{R}{0.4\linewidth}
% \vspace{-3mm}
\centering
\caption{Ablation study of intent identification (\texttt{IID}) on the \texttt{live-score} benchmark.}
\vspace{-2mm}
\label{tab:ec_method_comparison}
\resizebox{0.99\linewidth}{!}{%
\begin{tabular}{l|ccc}
\toprule
\textbf{Method} & \textbf{Pass@3} & \textbf{Tokens (M)} & \textbf{Steps} \\
\midrule
baseline     & 0.57 & 0.70 & 22.4  \\
\midrule
\rowcolor{gray!15}
\pjn{}-NL                & \textbf{0.93} & \textbf{0.39} & \textbf{17.3}  \\
\pjn{}-NL w/o \texttt{IID}     & 0.92 & 0.51 & 22.5  \\
\midrule
\rowcolor{gray!15}
\pjn{}              & \textbf{0.98} & \textbf{0.51} & \textbf{21.6} \\
\pjn{} w/o \texttt{IID}   & 0.98 & 0.66 & 28.1  \\
\bottomrule
\end{tabular}}
\end{wraptable}

\pjn{} has close to 100\% success rates, and our analysis of the failure cases shows that the remaining failures typically arise in \texttt{import} handling. 
In Terraform, \texttt{import} is used to bring externally created resources under IaC management by linking them with corresponding configuration blocks.
In one case from the \texttt{flask} project, when importing a resource, the agent mistakenly attempted \texttt{``RootAccUsage:/aws/cloudtrail/flask-micros''}, which incorrectly reversed the name and identifier, and consequently failed to discover the correct format. The correct format is 
\texttt{``/aws/cloudtrail/flask-micros:RootAccUsage.''}
% \begin{center}
% \footnotesize
% \begin{BVerbatim}
% import {
%     to = aws_cloudwatch_log_metric_filter.RootAccountUsage
%     id = “/aws/cloudtrail/flask-micros:RootAccountUsage”
% }
% \end{BVerbatim}
% \end{center}
Such cases could be addressed with retrieval-augmented generation (RAG), where the precise import syntax is supplied directly to the agent.
In summary, these results demonstrate that \pjn{} yields substantial accuracy and efficiency gains.

\subsection{RQ3: Intent Identification} 
% \ang{Confused by "annotation" and "consolidation" - these are two different techs we discussed in the design, but here are they used interchangeably?} 
\label{sec:ec_ablation}
\begin{figure}[t]
\centering
\begin{subfigure}{0.32\textwidth}
    \centering
    \includegraphics[width=\textwidth]{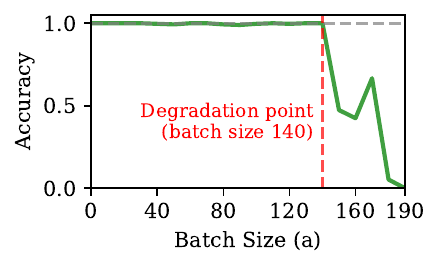}
\end{subfigure}
\begin{subfigure}{0.32\textwidth}
    \centering
    \includegraphics[width=\textwidth]{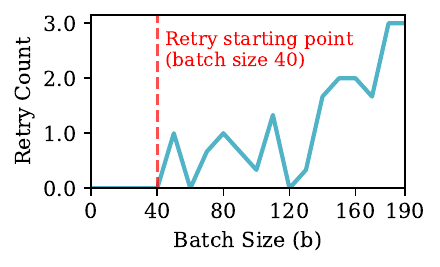}
\end{subfigure}
\begin{subfigure}{0.32\textwidth}
    \centering
    \includegraphics[width=\textwidth]{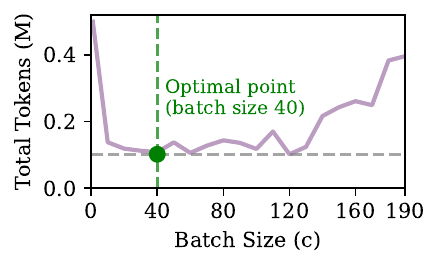}
\end{subfigure}
\vspace{-2mm}
\caption{LLM annotation of a synthetic API trace with 190 mutating events.}
\label{fig:batch_annotation_figures}
\end{figure}

\looseness=-1
Next (RQ3), we show that intent identification improves efficiency while maintaining accuracy. 
The small annotation overhead from LLM is amortized by the much larger savings during reconciliation.
Table~\ref{tab:ec_method_comparison} compares \pjn{} with and without intent identification (\texttt{IID}) on the \texttt{live-score} benchmark. 
Both \pjn{} and \pjn{}-NL maintain high accuracy, while \texttt{IID} increases their efficiency by 23\% in token usage and steps. Accuracy remains unchanged for \pjn{} (0.98) and even slightly improves for \pjn{}-NL (0.93 vs.\ 0.92).

Intent identification proceeds in two steps: annotation and consolidation. 
We next examine the scalability of its annotation stage—specifically, whether LLMs can annotate large batches of API events effectively.
We find that LLM annotation consumes on average 5.6K tokens per drift—only about 1\% of the overall reconciliation cost, which is on the order of millions of tokens.
% We next study the annotation mechanism in more detail, asking whether API events can be annotated effectively in batches. 
To stress-test the system, we construct a single trace containing 190 mutating API calls. 
Figure~\ref{fig:batch_annotation_figures} shows the effect of varying batch size on this trace. 
Accuracy drops sharply beyond 140, marking a clear degradation point. 
The number of retries increases once the batch size exceeds 40, indicating that the model struggles to process overly large batches in a single pass. This effect is also visible in token usage: it is minimized around a batch size of 40, which we identify as the optimal operating point. Beyond this point, additional retries required to handle larger batches lead to higher overall token consumption.
These results show that event consolidation is essential, and that batching around 40 mutating API calls is sufficient in practice, providing a good balance of accuracy, efficiency, and robustness. We therefore adopt a batch size of 40 for all experiments.

\subsection{RQ4: Patch Generation}

\begin{wraptable}{R}{0.4\textwidth}
% \vspace{-3mm}
\centering
\caption{Performance ablation study on the \texttt{live-score} benchmark.}
\vspace{-2mm}
\label{tab:method_comparison}
\resizebox{0.99\linewidth}{!}{%
\begin{tabular}{l|c}
\toprule
\textbf{Method} & \textbf{Pass@3} \\
\midrule
\rowcolor{gray!15}
\pjn{}          & \textbf{0.98}  \\
% \rowcolor{gray!15}
\pjn{}-NL         & 0.93  \\
\pjn{} w/o \texttt{drift\_report} & 0.60 \\
\pjn{} w/o \texttt{self\_critique} & 0.81  \\
\bottomrule
\end{tabular}
}
% \vspace{-2mm}
\end{wraptable}

\begin{figure}[t]
\centering
\begin{subfigure}{0.45\textwidth}
    \centering
    \includegraphics[width=\textwidth]{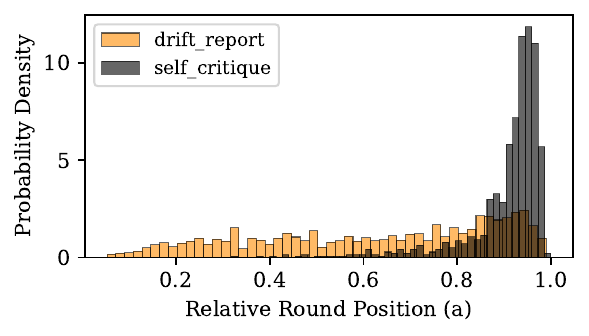}
    % \caption{\pjn{} without learning}
    % \label{fig:tool_timef_tools}
\end{subfigure}
% \hfill
\begin{subfigure}{0.45\textwidth}
    \centering
    \includegraphics[width=\textwidth]{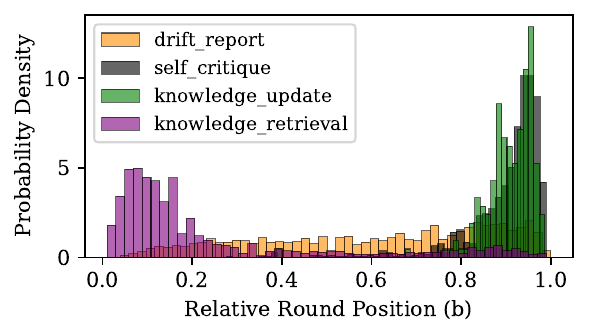}
    % \caption{\pjn{}}
    % \label{fig:tool_timef_leanring}
\end{subfigure}
    \vspace{-2mm}
    \caption{Distribution of tool usage across reconciliation progress. 
The x-axis denotes the normalized round position (\texttt{tooluse\_round} / \texttt{total\_round}) for each run. 
Figure~\ref{fig:tool_use_time}a In \pjn{}-NL, \texttt{drift\_report} is used uniformly throughout the process, while \texttt{self\_critique} is concentrated near the end. 
Figure~\ref{fig:tool_use_time}b In \pjn{}, similar patterns are observed, but knowledge tools exhibit distinct timing: \texttt{knowledge\_retrieval} is typically invoked early to supply prior context, while \texttt{knowledge\_update} appears near the end to record new insights.}
\label{fig:tool_use_time}
\end{figure}

%To answer RQ4, 
Table~\ref{tab:method_comparison} shows that specialized tools substantially improve reconciliation performance, with each removal lowering accuracy and the absence of \texttt{drift\_report} causing the steepest drop.
We perform an ablation study on the \texttt{live-score} benchmark. 
Removing any tool or knowledge-base (KB) operation reduces reconciliation accuracy. 
Without learning---i.e., in \pjn{}-NL, where the KB operations \texttt{knowledge\_update} and \texttt{knowledge\_retrieval} are disabled---accuracy drops from 0.98 to 0.93. 
Excluding \texttt{self\_critique} reduces it further to 0.81, while excluding \texttt{drift\_report} has the most severe impact, lowering accuracy to 0.60.
% Removing any tool reduces reconciliation accuracy: excluding learning drops accuracy from 0.98 to 0.93, excluding \texttt{self\_critique} reduces it further to 0.81, and excluding \texttt{drift\_report} has the most severe effect, lowering accuracy to 0.60. 
This shows that each tool plays an important role, with \texttt{drift\_report} and \texttt{self\_critique} being especially critical for achieving high performance. 

We next analyze the timing of tool use during reconciliation. 
As shown in Figure~\ref{fig:tool_use_time}, \texttt{drift\_report} is invoked by the agent throughout the process, \texttt{self\_critique} is concentrated near the end, and in \pjn{}, knowledge tools follow a clear temporal structure (\texttt{retrieval} at earlier stages, \texttt{update} at later stages). 
\begin{wrapfigure}{R}{0.45\textwidth}
    % \vspace{-2mm}
    \centering
    \includegraphics[width=\linewidth]{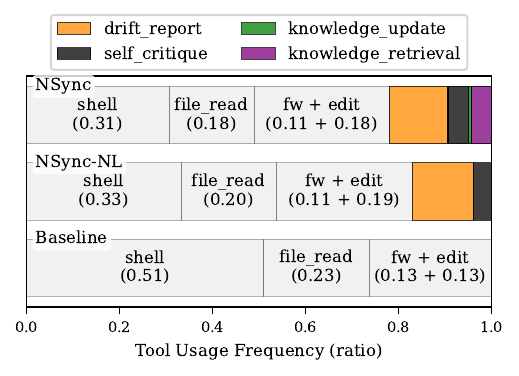}
    % \vspace{-3mm}
    \caption{Tool usage frequency breakdown.}
    \label{fig:toolusage_freq}
\end{wrapfigure}
This confirms that the agent is able to use our tools effectively, in a structured and phase-specific manner. 

\looseness=-1
Taking a closer look to tool usage frequency across the three methods (Figure~\ref{fig:toolusage_freq}). 
While the baseline relies heavily on generic \texttt{file\_read} and shell operations, \pjn{}-NL shifts its usage toward domain-specific tools, \texttt{drift\_report} and \texttt{self\_critique}. 
The frequency of editor and \texttt{file\_write} calls remains similar across agents, since these tools are still required to generate patches. 
By contrast, \texttt{shell} usage decreases substantially, as the baseline often used \texttt{shell} to locate files and invoke native Terraform commands—tasks that are largely replaced by domain-specific tools in \pjn{}. 
\texttt{file\_read} usage also decreases, since the baseline repeatedly re-scans the codebase to rediscover context, whereas \pjn{} makes more efficient use of retained context and KB. 
\pjn{} further incorporates KB operations, though at relatively low frequency compared to other tools; this is also reflected in Figure~\ref{fig:tool_use_time} as KB operation only happens at the beginning and the end.

\subsection{RQ5: Continual Learning}

\looseness=-1
To answer RQ5, we find that \pjn{} remain the most robust under stricter pass@1 evaluation when using our methods, as shown in Table~\ref{tab:pass_at_one_main_results}; it also maintains reliable performance across experiment runs where drifts are presented in arbitrary order. 
Figure~\ref{fig:kb_examples} presents a snippet of the evolving reconciliation knowledge base learned from experience. The knowledge spans diverse categories, including how to inspect resource state with Terraform commands, map API parameters to Terraform attributes, resolve formatting issues, and apply the custom tool \texttt{drift\_report} to pinpoint attribute-level drifts more effectively.

% \begin{table}[t]
% \centering
% \caption{Performance across runs: mean (min,max) $\pm$ std.}
% \vspace{-2mm}
% \resizebox{0.85\textwidth}{!}{%
% \begin{tabular}{r|cc|cc|cc}
% \toprule
% \multirow{2}{*}{\textbf{Benchmark}} & \multicolumn{2}{c}{\textbf{Baseline}} & \multicolumn{2}{c}{\textbf{\pjn{}}} & \multicolumn{2}{c}{\textbf{\pjn{}-L}} \\
% \cmidrule(lr){2-3} \cmidrule(lr){4-5} \cmidrule(lr){6-7}
%  & pass@1 & range $\pm$ std & pass@1 & range $\pm$ std & pass@1 & range $\pm$ std \\
% \midrule
% lab12 & 0.63 & (0.52, 0.78) $\pm$ 0.11 & 0.88 & (0.88, 0.89) $\pm$ 0.01 & \textbf{0.90} & (0.86, 0.95) $\pm$ 0.04 \\
% flask & 0.40 & (0.31, 0.47) $\pm$ 0.07 & \textbf{0.70} & (0.69, 0.72) $\pm$ 0.01 & 0.65 & (0.55, 0.74) $\pm$ 0.08 \\
% ssm3 & 0.51 & (0.49, 0.52) $\pm$ 0.01 & 0.73 & (0.65, 0.80) $\pm$ 0.06 & \textbf{0.80} & (0.76, 0.84) $\pm$ 0.03 \\
% live-score & 0.38 & (0.34, 0.39) $\pm$ 0.03 & 0.70 & (0.66, 0.75) $\pm$ 0.04 & \textbf{0.84} & (0.80, 0.93) $\pm$ 0.06 \\
% mega-mesh & 0.53 & (0.47, 0.59) $\pm$ 0.05 & 0.76 & (0.59, 0.88) $\pm$ 0.13 & \textbf{0.78} & (0.71, 0.82) $\pm$ 0.06 \\
% \bottomrule
% \end{tabular}}
% \end{table}

\begin{table}[t]
\centering
\caption{Performance across runs under the pass@1 metric}
\vspace{-2mm}
\resizebox{0.89\textwidth}{!}{%
\begin{tabular}{rcccccc}
\toprule
\multirow{2}{*}{\textbf{Benchmark}} & \multicolumn{2}{c}{\textbf{Baseline}} & \multicolumn{2}{c}{\textbf{\pjn{}-NL}} & \multicolumn{2}{c}{\textbf{\pjn{}}} \\
\cmidrule(lr){2-3} \cmidrule(lr){4-5} \cmidrule(lr){6-7}
 & Pass@1 & range $\pm$ std & Pass@1 & range $\pm$ std & Pass@1 & range $\pm$ std \\
\midrule
lab12 & 0.63 & (0.52, 0.78) $\pm$ 0.11 & 0.88 & (0.88, 0.89) $\pm$ 0.01 & \textbf{0.90} & (0.86, 0.95) $\pm$ 0.04 \\
flask & 0.40 & (0.31, 0.47) $\pm$ 0.07 & \textbf{0.70} & (0.69, 0.72) $\pm$ 0.01 & 0.65 & (0.55, 0.74) $\pm$ 0.08 \\
ssm3 & 0.51 & (0.49, 0.52) $\pm$ 0.01 & 0.73 & (0.65, 0.80) $\pm$ 0.06 & \textbf{0.80} & (0.76, 0.84) $\pm$ 0.03 \\
live-score & 0.38 & (0.34, 0.39) $\pm$ 0.03 & 0.70 & (0.66, 0.75) $\pm$ 0.04 & \textbf{0.84} & (0.80, 0.93) $\pm$ 0.06 \\
mega-mesh & 0.53 & (0.47, 0.59) $\pm$ 0.05 & 0.76 & (0.59, 0.88) $\pm$ 0.13 & \textbf{0.78} & (0.71, 0.82) $\pm$ 0.06 \\
\midrule
\textbf{Overall} & 0.49 & (0.43, 0.55) $\pm$ 0.05 & 0.76 & (0.69, 0.81) $\pm$ 0.05 & \textbf{0.80} & (0.73, 0.86) $\pm$ 0.05 \\
\bottomrule
\end{tabular}}
\label{tab:pass_at_one_main_results}
\end{table}

Table~\ref{tab:pass_at_one_main_results} evaluates robustness under the pass@1 metric, where no retries are allowed and results are averaged across three runs. The baseline performs poorly in this setting, with an overall accuracy of only 0.49, highlighting its unreliability when a single attempt must succeed. \pjn{}-NL improves substantially to 0.76, and \pjn{} further increases robustness to 0.80 by reusing knowledge from prior reconciliations. Importantly, the drift scenarios are presented in arbitrary order across runs, ensuring that improvements are not due to any favorable ordering of the data. This confirms that the learning method is agnostic to ordering and that there is no hidden bias or structure in our benchmarks that guarantees improvement. While all methods see a drop compared to pass@3, the degradation is smaller for \pjn{}-NL and especially for \pjn{}, which remains consistently strong across runs. These results demonstrate that learning not only improves average accuracy but also reduces variance, yielding more reliable reconciliation outcomes.  

\begin{figure}[t]
\centering
\begin{subfigure}{0.45\textwidth}
    \centering
    \includegraphics[width=\textwidth]{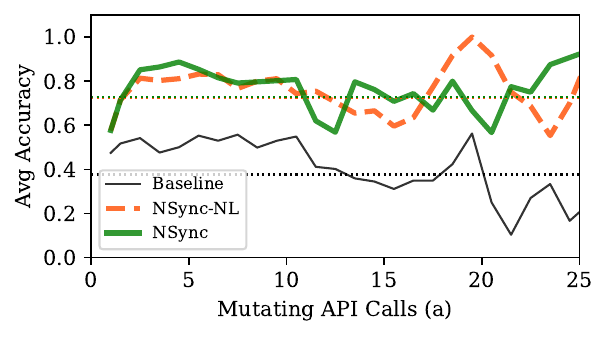}
    % \caption{}
    % \label{fig:acc-scale}
\end{subfigure}
% \hfill
\begin{subfigure}{0.45\textwidth}
    \centering
    \includegraphics[width=\textwidth]{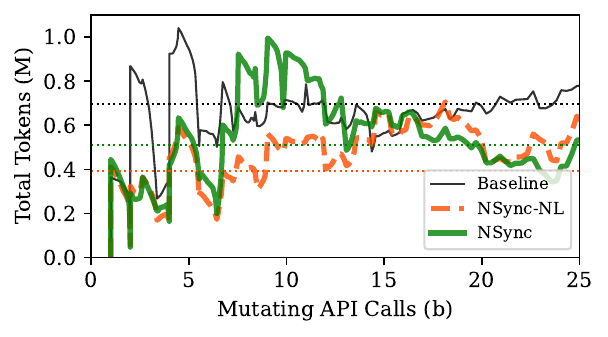}
    % \caption{}
    % \label{fig:token-scale}
\end{subfigure}
\vspace{-2mm}
\caption{Performance comparison across mutating API calls. Figure~\ref{fig:main_scaling}a reports average accuracy across all drift scenarios from five projects, reorganized by the number of mutating API calls. Figure~\ref{fig:main_scaling}b presents the corresponding token cost for reconciliation as drift complexity increases.}
% \vspace{-2mm}
\label{fig:main_scaling}
\end{figure}

\begin{figure}[t]
\centering
\resizebox{0.95\textwidth}{!}{%
\begin{tcolorbox}[colback=purple!5!white,
                  colframe=purple!40!black,
                  coltitle=white,
                  fonttitle=\bfseries,
                  title=A snippet of the evolving reconciliation knowledge base for \texttt{lab12}.,
                  before upper={%
                    \raggedright
                    \setlist[itemize]{leftmargin=0pt,
                                      itemsep=0pt,topsep=0pt,parsep=0pt,partopsep=0pt}%
                    \renewcommand\labelitemi{\textbullet}}]
\begin{itemize}\small
  \item When working with VPC flow logs, remember to escape the \texttt{log\_format} string with double dollar signs (\texttt{\$\$}) to avoid Terraform interpolation errors.
  \item Use \texttt{terraform state show <resource\_name>} to examine the current state of a specific resource.
  \item Use \texttt{moved} blocks in \texttt{move.tf} to handle resource renaming and prevent unnecessary resource recreation.
  \item For EventBridge rules, use \texttt{is\_enabled = false} to disable rules instead of \texttt{state = ``DISABLED''} which is the API parameter but not the Terraform attribute name
  \item For CloudTrail resources, ensure the S3 bucket policy includes the correct permissions for CloudTrail to write logs.
  \item When importing KMS keys, remove \texttt{deletion\_window\_in\_days} if it causes drift.
  \item When importing SSM documents, ensure the content format (quotes, indentation, etc.) exactly matches the actual state.
  \item When importing S3 buckets created outside of Terraform, make sure to also import any associated bucket policies separately.
  \item Use \texttt{drift\_report} to identify specific attributes that need to be changed in resources.
\end{itemize}
\end{tcolorbox}
}
\vspace{-3mm}
\caption{An example of the evolving reconciliation knowledge base. All entries for the \texttt{lab12} project represent accumulated knowledge, continually updated by the agent from its prior reconciliation experience}
% \vspace{-2mm}
\label{fig:kb_examples}
\end{figure}

Figure~\ref{fig:main_scaling} shows that, as drift complexity increases (more mutating API calls), the baseline shows a clear downward trend in accuracy, while \pjn{}-NL remain stable. 
In terms of efficiency, the baseline’s token usage steadily grows with complexity, whereas \pjn{}-NL consistently requires fewer tokens, and \pjn{} adds only a modest overhead due to knowledge retrieval. 
These results highlight that our methods are more resilient under increasing drift complexity, maintaining accuracy and efficiency where the baseline degrades. 
when numbers of mutating API increases the baseline token usage eventually suppress \pjn{}.

\section{Discussion}
\label{sec:discussion}

% Terraformer, aws2tf, Terracognita, aztfexport, and others~\cite{terraformer, terraformimport, aws2tf, terraform, aztfexport, gcp_tf_export, terracognita_2025} 
% \looseness=-1
\textbf{IaC lifting.} 
A class of IaC tools such as \texttt{Terraform-cfg-gen}~\cite{hashicorp2024generating}, Terraformer~\cite{terraformer}, aws2tf~\cite{aws2tf},  Terracognita~\cite{terracognita_2025}, aztfexport~\cite{aztfexport} attempt a related but different goal---to ``lift'' a non-IaC managed infrastructure to an IaC configuration from scratch. This need arises because some companies want to port their legacy infrastructure (e.g., created via API scripts) to be managed by IaC. This requires reconstructing configurations by issuing these APIs calls to retrieve the fragmented cloud state, and then inferring IaC-level configurations using heuristic mappings. Lifting attempts at a different goal from reconciliation, and is inherently more difficult: these tools add support for each new resource or service manually, and continuously track updates from both the cloud provider side (new APIs, changing semantics) and the IaC side (schema and language evolution). In practice, these tools produce brittle code with limited coverage~\cite{lilac}, embedding raw runtime values (e.g., identifiers, subnet IDs) instead of reusable references. The result is flat, non-modular configurations with generic names that are neither maintainable nor reusable. And importantly, they aim at lifting from an infrastructure to new configurations from scratch, rather than producing targeted patches to an existing IaC project. For larger infrastructures, these lifting tools degrade further, often producing syntax errors or unusable configurations~\cite{lilac}.  Simply put, these tools attempt at program synthesis from scratch, whereas ours aims at program repair. By reducing the complexity of the task, and by leveraging LLMs, \pjn{} sidesteps the key challenges that lead to brittleness in lifting tools.

\noindent \textbf{Beyond Terraform and AWS.}
Our evaluation is on the leading IaC platform---Terraform---and the most popular cloud---AWS. We believe that \pjn{}'s architecture is generalizable to other IaC frameworks (e.g., Pulumi~\cite{pulumi}, OpenTofu~\cite{opentofu}) and cloud providers (e.g., Azure, GCP). Key changes may include different API monitoring tools across clouds and resource management practices. For instance, Azure and GCP's event logs may use different event schemas compared to AWS CloudTrail. Each IaC framework also has its own local state scheme and planning operations. 
%Furthermore, the baseline comparison focuses on general-purpose LLM agents rather than existing infrastructure lifting tools. This decision was made because current lifting tools are not designed for reconciliation tasks and cannot operate over API traces. Nevertheless, future work should explore hybrid approaches that combine \pjn{}'s techniques with existing lifting capabilities. \ang{let's discuss.}
Hence, an interesting avenue of future work is to generalize \pjn{} to other IaC frameworks and clouds.

\section{Related Work}
\noindent \textbf{AI Agent for Cloud Management.}
Recent advances have explored the use of AI agents to automate diverse aspects of cloud operations, or ``AIOps.'' However, much of this work only focuses on  incident detection, root cause diagnosis, and remediation~\cite{AIOpsLab_mlsys25, Flow_of_Action_www25, RCACopilot_eurosys24, RCAgent_CIKM24, terrafault}---e.g., via log mining, causal inference, or LLM-powered analysis to recommend or  execute corrective actions. At the same time, several vision papers highlight the potential of AI agents to expand beyond diagnostic tasks to handle provisioning, continuous monitoring, and workload optimization~\cite{aiops19, aiops_vision23, cloudagent_vision}. \pjn{} is an AIOps design but is specialize to detecting and reconciling drift in IaC frameworks.

\vspace{.1cm}\noindent \textbf{AI Agent for Program Repair.}
Automated Program Repair (APR) has been extensively studied, with recent surveys showing that large language models (LLMs) have transformed the field by enabling repair through fine-tuning, prompting, procedural workflows, and agentic frameworks~\cite{yang2025survey, zhang2024survey}. 
Fine-tuned repair systems like VulMaster~\cite{zhou2024out}, RepairLLaMA~\cite{repairllama2023}, and RepairCAT~\cite{jiang2024repaircat} leverage domain-specific bug corpora for high accuracy, while prompting approaches such as AlphaRepair~\cite{xia2023alpharepair} and TracePrompt~\cite{haque2025towards} achieve lightweight deployment through zero/few-shot queries. 
More advanced pipelines integrate retrieval or analysis feedback to guide iterative repair (e.g., Repilot~\cite{wei2023copiloting}, Agentless~\cite{xia2024agentless}), and fully agentic systems such as SWE-Agent~\cite{yang2024swe}, RepairAgent~\cite{bouzenia2024repairagent}, OpenHands~\cite{wang2024openhands}, and AutoCodeRover~\cite{zhang2024autocoderover} orchestrate external tools, testing, and multi-step reasoning under LLM control. These works primarily target source code bugs in conventional software projects, with correctness validated against unit tests or benchmarks.
\pjn{} builds on the program repair perspective but in a novel domain: Infrastructure-as-Code. Unlike traditional APR, where explicit failing tests provide repair oracles, IaC reconciliation operates without runnable test cases: its ``specification'' is implicitly encoded in cloud API traces. This shifts the repair challenge from patching faulty logic based on test cases to synthesizing declarative updates that represent API traces, ensuring that the IaC codebase reflects the true deployed state.

\section{Conclusion} 
\label{sec:conclusion}

\looseness=-1
Infrastructure-as-Code (IaC) frameworks are gaining popularity, and they encode the infrastructure in a declarative configuration as the source of truth. However, when IaC frameworks are used together with other management interfaces (e.g., cloud consoles, CLI, SDK), the IaC configuration no longer captures the de facto state, leading to infrastructure drift. 
We presented \pjn{}, the first agentic system for automated IaC reconciliation. \pjn{} relies on cloud APIs to identify and reconcile drift, and it works by cleaning and consolidating noisy API traces, generating patches to update configurations, and continuously learning from experience. 
To support systematic evaluation, we introduced a pipeline for generating realistic and assessable drift scenarios, and curated the first IaC reconciliation dataset containing 372 validated cases across five real-world Terraform projects.
% Experiments show that \pjn{} achieves 0.97 accuracy—outperforming the baseline by 26\%—while using 22\% fewer tokens, and requires no human intervention. 
Experiments show that \pjn{} achieves 0.97 accuracy; a 26\% improvement over the baseline, while reducing token usage by 1.5$\times$.
These results highlight that IaC reconciliation is both feasible and practical with modern agentic systems.

\bibliographystyle{ACM-Reference-Format}
\bibliography{bib}

%%
%% If your work has an appendix, this is the place to put it.
% \appendix
\newpage 
\appendix 
\section{Benchmark Curation} 
\label{sec:appendix-bench}

This section details the benchmark curation process in Section~\ref{sec:eval_benchmark}. Our benchmark includes five Terraform projects and their associated mutations, designed to evaluate IaC reconciliation approaches. The five Terrafrom projects are from GitHub and listed in Table~\ref{tab:benchmarks}.

We create the benchmark to enable the following evaluation workflow. First, we deploy the original IaC configuration (one of the five Terraform projects) to provision the initial cloud infrastructure. Then, we inject infrastructure drift by applying one of our curated mutations to the same infrastructure. During this process, we collect two key components: (1) the API call traces that capture the drift-inducing operations, which serve as input to the reconciliation approach, and (2) the Terraform state from the mutated configuration, which serves as ground truth representing the up-to-date infrastructure state. To evaluate the correctness of a reconciliation approach's output, we execute \texttt{Terraform Plan} using the reconciled IaC configuration against this ground truth state. A successful reconciliation should result in no planned changes, indicating that the updated configuration accurately captures the infrastructure's current state.

For a IaC project, its mutations are generated via a LLM agent based on drift scenarios specified in natural language. We derived drift scenarios from two complementary sources: 
(1) \textbf{AWS Systems Manager (SSM) Automation runbooks}~\cite{aws2024runbook} and (2) \textbf{manually crafted drift scenarios}. The primary source, SSM Automation runbooks, are AWS-provided operational documents that contain both natural language descriptions and executable API scripts for common infrastructure management tasks. These runbooks represent real-world infrastructure modifications. We use their descriptions as drift scenarios. For each IaC project, we prompted an LLM with the runbook's drift scenario descriptions to generate corresponding mutated IaC configurations. We then validated each generated mutation through deployment testing - verifying that the mutation could be successfully deployed and reverted back to the original configuration.
We supplement the drift scenarios with a few manually-curated ones tailored to each IaC project. 

To complement our evaluation with persistent drifts, we also incorporated false-positive cases. These cases were created by first deploying a mutated configuration and then reverting it to the original state, simulating situations where infrastructure changes are initiated but subsequently rolled back. Although these cases generate API call traces that indicate infrastructure modifications occurred, they result in no persistent drift in the final infrastructure state. This approach allows us to test how well reconciliation tools handle traces that introduce non-persistent drifts. We sampled these false-positive cases from our overall mutation pool across different projects.

In total, we collected \textbf{372 mutation cases}, of which \textbf{127 are false-positive cases}. 
The following subsections describe the AWS runbook–based mutations and the manually crafted ones in detail.

\subsection{Sampling from AWS Automation Runbooks} 
\label{sec:appendix-runbook}

We began with a collection of 763 publicly available AWS Systems Manager (SSM) runbooks, which are pre-built operational scripts. Through a systematic filtering process, we narrowed these down to 90 unique scenarios (shown in Table~\ref{tab:runbook}). The filtering process excluded runbooks that do not perform infrastructure modifications, meaning they don't issue mutating API calls. We also removed runbooks that require pre-existing resources not included in our IaC project configurations. Additionally, we excluded runbooks that create irreversible or non-importable infrastructure changes, such as those that scale up EBS volumes (which cannot be scaled down) or create EBS volume images (which cannot be imported back into IaC). 

After filtering down to 90 unique runbook scenarios, we selected applicable scenarios for each IaC project based on their specific cloud resources. The selection is done using a LLM-based classifier. 
The classifier takes as input the runbook description and the resources provisioned by a Terraform project, and output whether the runbook can be meaningfully applied to mutate the AWS infrastructure. 
The prompt to the LLM is shown in Listing~\ref{lst:scenario-prompt}. 
The runbooks per project are listed in 
Tables~\ref{tab:per_ssm_csv_runbooks}, \ref{tab:per_lab12_csv_runbooks}, \ref{tab:per_flask_csv_runbooks},  \ref{tab:per_live-score_csv_runbooks}, and \ref{tab:per_mega_mesh_csv_runbooks}

For each applicable scenario, we employed an LLM to generate infrastructure mutations by analyzing both the IaC configuration and the scenario description, as illustrated in Figure~\ref{fig:dataset_collection}. To ensure the quality and reliability of these mutations, we implemented a two-step validation process: first verifying that each mutation could be successfully deployed, and then confirming that the infrastructure could be safely reverted to its original configuration after deployment. Only mutations that passed both validation steps were included in evaluation.

\clearpage
\pagebreak 

% \begin{table}[!ht]
% \centering
% \caption{AWS Runbook Scenarios for Infrastructure Mutations}
% \label{tab:runbook}
% \end{table}

\begin{scriptsize}
\begin{longtable}{cp{4.5cm}p{8cm}}
\caption{AWS Runbook Scenarios for Infrastructure Mutations\label{tab:runbook}}\vspace{-3mm}\\
\toprule
\textbf{No.} & \textbf{AWS Runbook} & \textbf{Description} \\
\midrule
\endfirsthead
\toprule
\textbf{No.} & \textbf{AWS Runbook} & \textbf{Description} \\
\midrule
\endhead
\bottomrule
\endfoot
\bottomrule
\endlastfoot
1 & AWS-EnableNeptuneDbBackupRetentionPeriod & This runbook enables automated backups for an Amazon Neptune DB cluster by setting a backup retention period between 7 and 35 days. \\ \addlinespace
2 & AWS-EnableVPCFlowLogs & This runbook creates VPC flow logs to capture IP traffic information for specified VPCs, publishing the logs to either CloudWatch or S3 for network traffic monitoring and analysis. \\ \addlinespace
3 & AWSConfigRemediation-DropInvalidHeadersForALB & This runbook configures an Application Load Balancer to remove HTTP headers with invalid fields by enabling the drop invalid headers attribute. \\ \addlinespace
4 & AWSSupport-ConfigureDNSQueryLogging & This runbook automates the configuration of DNS query logging for AWS Route 53 public hosted zones or VPCs, capturing DNS queries and publishing them to CloudWatch Logs, S3, or Kinesis to help with troubleshooting connectivity issues. \\ \addlinespace
5 & AWSDocs-S3StaticWebsiteCustomDomain & This runbook automates the configuration of a static website hosted on Amazon S3 using a custom domain registered with Route 53, creating necessary S3 buckets, configuring website hosting, and setting up domain routing. \\ \addlinespace
6 & AWSConfigRemediation-EnableSecurityHub & This runbook enables AWS Security Hub for an AWS account in the current region and verifies its enabled status. \\ \addlinespace
7 & AWSConfigRemediation-EnablePITRForDynamoDbTable & This runbook enables Point-In-Time Recovery (PITR) for a specified Amazon DynamoDB table and verifies the feature was successfully activated. \\ \addlinespace
8 & AWSFleetManager-AddUsersToGroups & This runbook adds a list of specified users to multiple groups across Windows and Linux systems, verifying user and group existence before attempting the additions. \\ \addlinespace
9 & AWSEC2Launch-RunMigration & This runbook automates the migration process from EC2Config and EC2Launch v1 to EC2Launch v2 on Windows instances, with an option to perform a dry run before actual migration. \\ \addlinespace
10 & AWS-RestrictIncomingTraffic & This runbook restricts incoming TCP traffic to EC2 security groups by removing ingress rules that allow unrestricted access (0.0.0.0/0 and ::/0) on specified ports. \\ \addlinespace
11 & AWS-RemediateSSMAgentVPCEndpoints & This runbook creates VPC endpoints required by SSM Agent if they don't exist, and associates them with one subnet in each availability zone to ensure proper SSM connectivity. \\ \addlinespace
12 & AWS-CreateManagedLinuxInstance & This runbook automates the creation of a Linux managed instance in AWS with Systems Manager (SSM) integration, configuring necessary components like security groups, IAM roles, and the SSM agent for remote management. \\ \addlinespace
13 & AWS-SetRequiredTags & This runbook adds specified tags to one or more AWS resources across various services, with the ability to track both successful and failed tagging operations. \\ \addlinespace
14 & AWSConfigRemediation-CreateGuardDutyDetector & This automation enables Amazon GuardDuty by creating and activating a detector in the current AWS region. \\ \addlinespace
15 & AWS-EnableStepFunctionsStateMachineLogging & This runbook enables or updates logging configuration on an AWS Step Functions State Machine, specifying logging levels, CloudWatch log group destination, execution data inclusion options, and X-Ray tracing. \\ \addlinespace
16 & AWS-EnableCloudTrail & This runbook creates and enables a new CloudTrail trail to log AWS API activity, directing the log files to a specified S3 bucket. \\ \addlinespace
17 & AWS-EnableCWAlarm & This runbook creates CloudWatch alarms for AWS resources (EC2 instances, EBS volumes, S3 buckets, and RDS clusters) that don't already have them, configuring metric-based monitoring with customizable thresholds and comparison operators. \\ \addlinespace
18 & AWS-ArchiveS3BucketToIntelligentTiering & This runbook creates or modifies an S3 Intelligent-Tiering configuration to automatically optimize storage costs by moving data to more cost-effective tiers based on access patterns. \\ \addlinespace
19 & AWSConfigRemediation-CreateCloudTrailMultiRegionTrail & This runbook creates a multi-region CloudTrail trail that logs activities across multiple AWS regions and delivers the log files to a specified S3 bucket with log file validation enabled. \\ \addlinespace
20 & AWS-EnableS3BucketEncryption & This runbook enables default server-side encryption on an Amazon S3 bucket using the specified encryption algorithm (defaulting to AES256). \\ \addlinespace
21 & AWS-CreateSnapshot & The runbook creates a snapshot of a specified EBS volume and waits for the snapshot creation process to complete. \\ \addlinespace
22 & AWSConfigRemediation-DetachIAMPolicy & This runbook detaches a specified IAM policy from all entities (groups, users, and roles) it is attached to, verifying complete removal after execution. \\ \addlinespace
23 & AWSConfigRemediation-EnableLoggingForALBAndCLB & This runbook enables and verifies logging for AWS Application Load Balancers (ALB) or Classic Load Balancers (CLB) by configuring them to send access logs to a specified S3 bucket. \\ \addlinespace
24 & AWSConfigRemediation-UpdateXRayKMSKey & This runbook enables encryption for AWS X-Ray data using a customer-managed AWS KMS key to meet security best practices. \\ \addlinespace
25 & AWSConformancePacks-SecurityBestPracticesforAWSWAF & This runbook deploys AWS Config rules to enforce security best practices for AWS WAF, ensuring WAF is enabled on Application Load Balancers and API Gateway stages, and verifying that WAF Regional rules, rule groups, and web ACLs are properly configured and not empty. \\ \addlinespace
26 & AWSSupport-ConfigureEC2Metadata & This runbook helps configure EC2 Instance Metadata Service (IMDS) options, allowing users to enforce IMDSv2, set the HTTP hop limit, or disable metadata access entirely.

Note: Changes to IMDS configuration should be made cautiously as they may break applications that rely on specific metadata service versions or access. \\ \addlinespace
27 & AWSConfigRemediation-EnableVPCFlowLogsToS3Bucket & This runbook replaces an existing VPC Flow Log that publishes to CloudWatch Logs with one that publishes to an Amazon S3 bucket instead. \\ \addlinespace
28 & AWS-EnableCloudTrailCloudWatchLogs & This runbook configures CloudTrail trails to deliver their events to a specified CloudWatch log group, enabling log monitoring and analysis capabilities. \\ \addlinespace
29 & AWSConformancePacks-SecurityBestPracticesforCloudTrail & This runbook implements security best practices for CloudTrail by configuring AWS Config rules that monitor and enforce CloudTrail settings like encryption, log validation, multi-region logging, CloudWatch Logs integration, and S3 data event tracking. \\ \addlinespace
30 & AWSConfigRemediation-ConfigureS3PublicAccessBlock & This runbook creates or modifies the S3 PublicAccessBlock configuration for an AWS account to restrict public access to S3 buckets. \\ \addlinespace
31 & AWS-AttachIAMToInstance & This runbook attaches an IAM role to an EC2 instance, with the ability to create a new instance profile if needed and optionally replace an existing IAM profile. \\ \addlinespace
32 & AWS-DisableEventBridgeRule & This runbook disables a specified rule in Amazon EventBridge, allowing users to temporarily prevent a rule from triggering events. \\ \addlinespace
33 & AWSConfigRemediation-EnableAutoScalingGroupELBHealthCheck & This runbook enables Elastic Load Balancer (ELB) health checks for an Amazon EC2 Auto Scaling group and sets an appropriate grace period before those health checks begin. \\ \addlinespace
34 & AWSSupport-ConfigureTrafficMirroring & This runbook configures traffic mirroring to help troubleshoot connectivity issues between load balancers and EC2 instances by copying inbound and outbound network traffic from network interfaces. \\ \addlinespace
35 & AWS-DisableS3BucketPublicReadWrite & This runbook disables public read and write access to an S3 bucket by applying public access block settings. \\ \addlinespace
36 & AWSSupport-SetupIPMonitoringFromVPC & This runbook sets up continuous network monitoring for specified IPv4/IPv6 addresses by creating an EC2 instance that runs ping, MTR, traceroute, and traceTCP tests at regular intervals, with results sent to CloudWatch Logs and visualized on a dashboard for troubleshooting network issues. \\ \addlinespace
37 & AWSConfigRemediation-ConfigureS3BucketPublicAccessBlock & This runbook creates or modifies the PublicAccessBlock configuration for an S3 bucket to control public access settings, helping to enhance security by restricting public access to bucket contents. \\ \addlinespace
38 & AWS-UpdateALBDesyncMitigationMode & This runbook updates the desync mitigation mode on an Application Load Balancer (ALB) to help manage how the load balancer handles requests that might pose security risks to applications. \\ \addlinespace
39 & AWSConfigRemediation-DeleteDefaultVPCRoutes & This runbook deletes default routes (0.0.0.0/0 and ::/0) from a specified Amazon EC2 VPC route table to enhance network security by removing unrestricted internet access. \\ \addlinespace
40 & AWSSupport-RemediateLambdaS3Event & This runbook troubleshoots and remedies issues with S3 event triggers for Lambda functions by checking event configurations, adding necessary permissions, and fixing resource policy problems. \\ \addlinespace
41 & AWSConfigRemediation-DeleteUnusedSecurityGroup & This runbook deletes non-default security groups that are not being utilized by elastic network interfaces, while preserving any security group named "default". \\ \addlinespace
42 & AWS-RemediateSSMAgentHTTPSAccess & This runbook configures EC2 security groups to enable HTTPS communication between SSM Agent on EC2 instances and Systems Manager through VPC endpoints. \\ \addlinespace
43 & AWSConfigRemediation-EnableAccountAccessAnalyzer & This runbook creates an IAM Access Analyzer at the AWS account level to help identify resources that are shared with external entities. \\ \addlinespace
44 & AWSConfigRemediation-SetIAMPasswordPolicy & This runbook sets and verifies the IAM password policy for an AWS account, configuring parameters such as password length, complexity requirements, expiration, and reuse prevention. \\ \addlinespace
45 & AWS-CreateS3PolicyToExpireMultipartUploads & Creates and applies an S3 bucket lifecycle policy to automatically expire incomplete multipart uploads after a specified number of days, while preserving any existing lifecycle rules. \\ \addlinespace
46 & AWS-RemediateSSMAgentVPCAttributes & This runbook remediates VPC attribute issues by enabling DNS support and DNS hostnames in VPCs to fix unmanaged EC2 instance connectivity problems. \\ \addlinespace
47 & AWS-EnableS3BucketKeys & This runbook enables S3 Bucket Keys on a specified S3 bucket to create data keys for new objects, supporting either server-side encryption with S3 managed keys (SSE-S3) or AWS KMS keys (SSE-KMS). \\ \addlinespace
48 & AWS-ConfigureDocker & This runbook configures Windows instances to either install or uninstall Docker and container functionality. \\ \addlinespace
49 & AWS-SetupManagedRoleOnEc2Instance & This runbook configures an EC2 instance with a managed IAM role that has the AmazonSSMManagedInstanceCore policy, enabling Systems Manager management capabilities for the instance. \\ \addlinespace
50 & AWS-DisablePublicAccessForSecurityGroup & This runbook disables SSH (port 22) and RDP (port 3389) access in a specified security group, either from all IP addresses or from a specific IPv4/IPv6 address. \\ \addlinespace
51 & AWS-AttachExcludeConditionToS3DenyPolicies & This runbook excludes a specified principal from all Deny statements in an S3 bucket's resource-based policies by adding appropriate ArnNotEquals conditions. \\ \addlinespace
52 & AWSConfigRemediation-EnforceEC2InstanceIMDSv2 & This runbook enforces IMDSv2 on an EC2 instance by modifying its metadata options to require token-based authentication, enhancing security against potential SSRF vulnerabilities. \\ \addlinespace
53 & AmazonCloudWatch-MigrateCloudWatchAgent & This runbook automates the migration from the SSM CloudWatch Plugin to the Amazon CloudWatch Agent on Windows systems by checking agent compatibility, disabling the old agent, installing the new agent, migrating existing configurations, and reconfiguring the new agent. \\ \addlinespace
54 & AWS-CreateDSManagementInstance & This runbook creates and configures a Windows management instance for AWS Directory Service, including domain joining the EC2 instance to your AWS-managed Active Directory and installing Active Directory administration tools. \\ \addlinespace
55 & AWSConfigRemediation-EnableCloudFrontOriginFailover & This runbook configures origin failover functionality for an Amazon CloudFront distribution by setting up an origin group that automatically redirects traffic to a secondary origin when the primary origin returns specified error codes. \\ \addlinespace
56 & AWS-JoinDirectoryServiceDomain & This runbook joins EC2 instances to a specified AWS Directory Service domain, allowing for centralized management of Windows instances through Active Directory. \\ \addlinespace
57 & AWSConfigRemediation-RemovePrincipalStarFromS3BucketPolicy & This runbook removes policy statements with wildcard principals (Principal: * or Principal: "AWS": *) for Allow actions from an S3 bucket policy to enhance security, deleting the entire bucket policy if it contains only such wildcard statements. \\ \addlinespace
58 & AWSConfigRemediation-EnableCloudFrontAccessLogs & This runbook enables and configures access logging for a specified Amazon CloudFront distribution, storing the logs in a designated S3 bucket with options for log file prefixing and cookie inclusion. \\ \addlinespace
59 & AWSConfigRemediation-EnableSystemsManagerSessionManager-AuditLogsToS3 & This runbook enables AWS Systems Manager Session Manager audit logging to an Amazon S3 bucket by creating or updating the SSM-SessionManagerRunShell document with the specified S3 bucket configuration. \\ \addlinespace
60 & AWSConfigRemediation-EnableNLBCrossZoneLoadBalancing & This runbook enables Cross Zone Load Balancing on an AWS Network Load Balancer and verifies that the setting was successfully applied. \\ \addlinespace
61 & AWSSupport-ContainS3Resource & This runbook isolates an S3 bucket or object in response to a security incident by restricting access policies, blocking public access, and enforcing bucket ownership controls, while preserving the resource for investigation. \\ \addlinespace
62 & AWSSupport-AnalyzeEBSResourceUsage & This runbook checks for unused EBS volumes, snapshots without source volumes, and unused AMIs, then generates CSV reports and stores them in a user-specified S3 bucket to help identify potential cost savings. \\ \addlinespace
63 & AWSConfigRemediation-DisableSubnetAutoAssignPublicIP & This runbook disables the automatic assignment of public IP addresses on a specified subnet by setting the MapPublicIpOnLaunch attribute to false. \\ \addlinespace
64 & AWS-ModifyDynamoDBProvisionedCapacity & This runbook modifies the read and write provisioned capacity of a DynamoDB table while handling table state transitions and considering auto-scaling configurations. \\ \addlinespace
65 & AWS-ConfigureS3BucketVersioning & This runbook enables or suspends versioning for a specified Amazon S3 bucket, configuring whether objects added to the bucket receive unique version IDs or null version IDs. \\ \addlinespace
66 & AWSConfigRemediation-EncryptSNSTopic & This runbook enables encryption on an Amazon SNS topic using a customer-managed KMS key to improve security compliance. \\ \addlinespace
67 & AWS-ConfigureCloudWatchOnEC2Instance & This runbook enables or disables CloudWatch monitoring on an EC2 instance. \\ \addlinespace
68 & AWSConformancePacks-OperationalBestPractices-forAmazonS3withRemediation & This runbook implements operational best practices for Amazon S3 by automatically detecting and remediating security issues including public access, encryption, and logging configurations for S3 buckets. \\ \addlinespace
69 & AWSConfigRemediation-EnableKeyRotation & This runbook enables automatic key rotation for an AWS KMS symmetric customer master key (CMK) and verifies that the rotation has been successfully enabled. \\ \addlinespace
70 & AWSDocs-S3StaticWebsite & This runbook automates the setup of a static website hosted on Amazon S3 by creating a publicly accessible S3 bucket with website hosting enabled and uploading index and error HTML documents. \\ \addlinespace
71 & AWS-EnableS3BucketEventNotifications & This runbook creates or updates Amazon S3 Event Notifications for a specified bucket to send alerts when selected events occur, helping detect accidental or intentional modifications that could lead to unauthorized data access. \\ \addlinespace
72 & AWSConfigRemediation-RestrictBucketSSLRequestsOnly & This runbook creates an S3 bucket policy that denies HTTP requests to a specified bucket, enforcing SSL/TLS encryption for all data transfers to enhance security. \\ \addlinespace
73 & AWS-ReleaseElasticIP & This runbook releases a specified Elastic IP address from an AWS account using its allocation ID. \\ \addlinespace
74 & AWS-ConfigureCloudTrailLogging & This runbook enables or disables logging for a specified CloudTrail based on the input parameter, first checking the current logging status to avoid redundant operations. \\ \addlinespace
75 & AWSSupport-EnableVPCFlowLogs & This runbook automates the creation of VPC flow logs for various AWS resources (subnets, elastic network interfaces, VPCs, transit gateways, and transit gateway attachments) with options to publish the logs to CloudWatch Logs or Amazon S3. \\ \addlinespace
76 & AWS-UpdateCLBDesyncMitigationMode & This runbook updates the desync mitigation mode on a Classic Load Balancer (CLB) to help protect applications from security risks posed by HTTP desynchronization attacks. \\ \addlinespace
77 & AWSSupport-SetupConfig & This runbook automates the setup of AWS Config, creating necessary components like service linked roles, recorder, S3 bucket, delivery channel, and optional aggregator authorizations if they don't already exist, otherwise leveraging existing resources. \\ \addlinespace
78 & AWSFleetManager-CreateUser & This runbook creates a local user account on both Windows and Linux systems, with options to specify a description and create a home directory on Linux. \\ \addlinespace
79 & AWS-DisableIncomingSSHOnPort22 & This runbook disables unrestricted incoming SSH traffic on port 22 for specified EC2 security groups by removing ingress rules that allow public access (from '0.0.0.0/0' and '::/0'). \\ \addlinespace
80 & AWSConfigRemediation-EnableCWLoggingForSessionManager & This runbook enables AWS Systems Manager Session Manager to store session output logs in a specified CloudWatch log group by creating or updating the necessary configuration document. \\ \addlinespace
81 & AWS-CreateEKSClusterWithFargateProfile & Creates an Amazon EKS cluster with a Fargate profile, providing a serverless Kubernetes environment where compute capacity is provisioned by AWS Fargate instead of EC2 instances. \\ \addlinespace
82 & AWS-SetupInventory & This runbook creates and manages an association between specified EC2 instances and a software inventory document, enabling automated collection of system inventory data on a scheduled basis. \\ \addlinespace
83 & AWSDocs-ScaleLoadBalanced & This runbook automates the setup of a scaled and load-balanced application by creating a launch template, auto scaling group, and target group, then associating the auto scaling group with a load balancer. \\ \addlinespace
84 & AWSConfigRemediation-EnableCloudTrailEncryptionWithKMS & This runbook encrypts an AWS CloudTrail trail using a specified AWS KMS customer master key to enhance security according to recommended best practices. \\ \addlinespace
85 & AWS-CloseSecurityGroup & This runbook closes a security group by removing all ingress and egress rules, effectively blocking all traffic to and from resources associated with the security group. \\ \addlinespace
86 & AWS-ConfigureS3BucketLogging & This runbook configures server access logging for an Amazon S3 bucket, directing log files to a target bucket with options for prefix configuration and permission settings. \\ \addlinespace
87 & AWSConfigRemediation-EncryptLambdaEnvironmentVariablesWithCMK & This runbook encrypts Lambda function environment variables at rest using a specified AWS KMS customer managed key to enhance security. \\ \addlinespace
88 & AWSConfigRemediation-RemoveUnrestrictedSourceIngressRules & This runbook removes all ingress rules from a specified security group that allow traffic from unrestricted source addresses (0.0.0.0/0 for IPv4 and ::/0 for IPv6) to improve security posture. \\ \addlinespace
89 & AWSDocs-LambdaWithS3SSMDocument & This runbook demonstrates an automated tutorial that sets up a Lambda function triggered by S3 events to create thumbnail images when new images are uploaded to a source bucket. \\ \addlinespace
90 & AWS-ChangeDDBRWCapacityMode & This runbook changes the read/write capacity mode for one or more DynamoDB tables, switching between on-demand mode and provisioned mode with appropriate capacity settings. \\ \addlinespace
\end{longtable}
\end{scriptsize}

\clearpage
\pagebreak 
% Requires: \usepackage{booktabs,array}
\begin{table}[t]
\centering
\caption{AWS Runbooks applicable to \texttt{ssm}}
\vspace{-3mm}
{\scriptsize
\begin{tabular}{ccp{9cm}}
\toprule
\textbf{No.} & \textbf{Idx} & \textbf{Scenario} \\
\midrule
1 & 69 & AWSConfigRemediation-EnableKeyRotation \\
2 & 10 & AWS-RestrictIncomingTraffic \\
3 & 55 & AWSConfigRemediation-EnableCloudFrontOriginFailover \\
4 & 71 & AWS-EnableS3BucketEventNotifications \\
5 & 34 & AWSSupport-ConfigureTrafficMirroring \\
6 & 26 & AWSSupport-ConfigureEC2Metadata \\
7 & 68 & AWSConformancePacks-OperationalBestPracticesforAmazonS3withRemediation \\
8 & 35 & AWS-DisableS3BucketPublicReadWrite \\
9 & 3 & AWSConfigRemediation-DropInvalidHeadersForALB \\
10 & 57 & AWSConfigRemediation-RemovePrincipalStarFromS3BucketPolicy \\
11 & 72 & AWSConfigRemediation-RestrictBucketSSLRequestsOnly \\
12 & 59 & AWSConfigRemediation-EnableSystemsManagerSessionManagerAuditLogsToS3 \\
13 & 18 & AWS-ArchiveS3BucketToIntelligentTiering \\
14 & 86 & AWS-ConfigureS3BucketLogging \\
15 & 24 & AWSConfigRemediation-UpdateXRayKMSKey \\
16 & 5 & AWSDocs-S3StaticWebsiteCustomDomain \\
17 & 25 & AWSConformancePacks-SecurityBestPracticesforAWSWAF \\
18 & 56 & AWS-JoinDirectoryServiceDomain \\
19 & 6 & AWSConfigRemediation-EnableSecurityHub \\
20 & 42 & AWS-RemediateSSMAgentHTTPSAccess \\
21 & 38 & AWS-UpdateALBDesyncMitigationMode \\
22 & 67 & AWS-ConfigureCloudWatchOnEC2Instance \\
23 & 89 & AWSDocs-LambdaWithS3SSMDocument \\
24 & 88 & AWSConfigRemediation-RemoveUnrestrictedSourceIngressRules \\
25 & 76 & AWS-UpdateCLBDesyncMitigationMode \\
26 & 23 & AWSConfigRemediation-EnableLoggingForALBAndCLB \\
27 & 45 & AWS-CreateS3PolicyToExpireMultipartUploads \\
28 & 28 & AWS-EnableCloudTrailCloudWatchLogs \\
29 & 50 & AWS-DisablePublicAccessForSecurityGroup \\
30 & 48 & AWS-ConfigureDocker \\
31 & 19 & AWSConfigRemediation-CreateCloudTrailMultiRegionTrail \\
32 & 61 & AWSSupport-ContainS3Resource \\
33 & 44 & AWSConfigRemediation-SetIAMPasswordPolicy \\
34 & 30 & AWSConfigRemediation-ConfigureS3PublicAccessBlock \\
35 & 51 & AWS-AttachExcludeConditionToS3DenyPolicies \\
36 & 74 & AWS-ConfigureCloudTrailLogging \\
37 & 43 & AWSConfigRemediation-EnableAccountAccessAnalyzer \\
38 & 14 & AWSConfigRemediation-CreateGuardDutyDetector \\
39 & 62 & AWSSupport-AnalyzeEBSResourceUsage \\
40 & 82 & AWS-SetupInventory \\
41 & 13 & AWS-SetRequiredTags \\
42 & 9 & AWSEC2Launch-RunMigration \\
43 & 22 & AWSConfigRemediation-DetachIAMPolicy \\
44 & 77 & AWSSupport-SetupConfig \\
45 & 66 & AWSConfigRemediation-EncryptSNSTopic \\
46 & 17 & AWS-EnableCWAlarm \\
47 & 52 & AWSConfigRemediation-EnforceEC2InstanceIMDSv2 \\
48 & 4 & AWSSupport-ConfigureDNSQueryLogging \\
49 & 83 & AWSDocs-ScaleLoadBalanced \\
50 & 60 & AWSConfigRemediation-EnableNLBCrossZoneLoadBalancing \\
51 & 16 & AWS-EnableCloudTrail \\
52 & 36 & AWSSupport-SetupIPMonitoringFromVPC \\
53 & 37 & AWSConfigRemediation-ConfigureS3BucketPublicAccessBlock \\
54 & 70 & AWSDocs-S3StaticWebsite \\
55 & 15 & AWS-EnableStepFunctionsStateMachineLogging \\
56 & 80 & AWSConfigRemediation-EnableCWLoggingForSessionManager \\
57 & 65 & AWS-ConfigureS3BucketVersioning \\
58 & 11 & AWS-RemediateSSMAgentVPCEndpoints \\
59 & 47 & AWS-EnableS3BucketKeys \\
60 & 29 & AWSConformancePacks-SecurityBestPracticesforCloudTrail \\
\bottomrule
\label{tab:per_ssm_csv_runbooks}
\end{tabular}
}
\end{table}

% \newpage 

% Requires: \usepackage{booktabs,array}
\begin{table}[t]
\centering
\caption{AWS Runbooks applicable to \texttt{lab12}}
\vspace{-3mm}
{\scriptsize
\begin{tabular}{ccp{9cm}}
\toprule
\textbf{No.} & \textbf{Idx} & \textbf{Scenario} \\
\midrule
1 & 10 & AWS-RestrictIncomingTraffic \\
2 & 31 & AWS-AttachIAMToInstance \\
3 & 68 & AWSConformancePacks-OperationalBestPracticesforAmazonS3withRemediation \\
4 & 81 & AWS-CreateEKSClusterWithFargateProfile \\
5 & 59 & AWSConfigRemediation-EnableSystemsManagerSessionManagerAuditLogsToS3 \\
6 & 8 & AWSFleetManager-AddUsersToGroups \\
7 & 6 & AWSConfigRemediation-EnableSecurityHub \\
8 & 21 & AWS-CreateSnapshot \\
9 & 42 & AWS-RemediateSSMAgentHTTPSAccess \\
10 & 27 & AWSConfigRemediation-EnableVPCFlowLogsToS3Bucket \\
11 & 88 & AWSConfigRemediation-RemoveUnrestrictedSourceIngressRules \\
12 & 28 & AWS-EnableCloudTrailCloudWatchLogs \\
13 & 50 & AWS-DisablePublicAccessForSecurityGroup \\
14 & 19 & AWSConfigRemediation-CreateCloudTrailMultiRegionTrail \\
15 & 44 & AWSConfigRemediation-SetIAMPasswordPolicy \\
16 & 30 & AWSConfigRemediation-ConfigureS3PublicAccessBlock \\
17 & 74 & AWS-ConfigureCloudTrailLogging \\
18 & 43 & AWSConfigRemediation-EnableAccountAccessAnalyzer \\
19 & 79 & AWS-DisableIncomingSSHOnPort22 \\
20 & 14 & AWSConfigRemediation-CreateGuardDutyDetector \\
21 & 53 & AmazonCloudWatch-MigrateCloudWatchAgent \\
22 & 62 & AWSSupport-AnalyzeEBSResourceUsage \\
23 & 82 & AWS-SetupInventory \\
24 & 39 & AWSConfigRemediation-DeleteDefaultVPCRoutes \\
25 & 13 & AWS-SetRequiredTags \\
26 & 49 & AWS-SetupManagedRoleOnEc2Instance \\
27 & 77 & AWSSupport-SetupConfig \\
28 & 17 & AWS-EnableCWAlarm \\
29 & 52 & AWSConfigRemediation-EnforceEC2InstanceIMDSv2 \\
30 & 4 & AWSSupport-ConfigureDNSQueryLogging \\
31 & 85 & AWS-CloseSecurityGroup \\
32 & 73 & AWS-ReleaseElasticIP \\
33 & 16 & AWS-EnableCloudTrail \\
34 & 36 & AWSSupport-SetupIPMonitoringFromVPC \\
35 & 70 & AWSDocs-S3StaticWebsite \\
36 & 63 & AWSConfigRemediation-DisableSubnetAutoAssignPublicIP \\
37 & 80 & AWSConfigRemediation-EnableCWLoggingForSessionManager \\
38 & 11 & AWS-RemediateSSMAgentVPCEndpoints \\
39 & 29 & AWSConformancePacks-SecurityBestPracticesforCloudTrail \\
\bottomrule
\label{tab:per_lab12_csv_runbooks}
\end{tabular}
}
\end{table}

\newpage 

% Requires: \usepackage{booktabs,array}
\begin{table}[t]
\centering
\caption{AWS Runbooks applicable to \texttt{flask}}
\vspace{-3mm}
{\scriptsize
\begin{tabular}{ccp{9cm}}
\toprule
\textbf{No.} & \textbf{Idx} & \textbf{Scenario} \\
\midrule
1 & 10 & AWS-RestrictIncomingTraffic \\
2 & 58 & AWSConfigRemediation-EnableCloudFrontAccessLogs \\
3 & 71 & AWS-EnableS3BucketEventNotifications \\
4 & 26 & AWSSupport-ConfigureEC2Metadata \\
5 & 3 & AWSConfigRemediation-DropInvalidHeadersForALB \\
6 & 57 & AWSConfigRemediation-RemovePrincipalStarFromS3BucketPolicy \\
7 & 72 & AWSConfigRemediation-RestrictBucketSSLRequestsOnly \\
8 & 59 & AWSConfigRemediation-EnableSystemsManagerSessionManagerAuditLogsToS3 \\
9 & 12 & AWS-CreateManagedLinuxInstance \\
10 & 18 & AWS-ArchiveS3BucketToIntelligentTiering \\
11 & 6 & AWSConfigRemediation-EnableSecurityHub \\
12 & 42 & AWS-RemediateSSMAgentHTTPSAccess \\
13 & 67 & AWS-ConfigureCloudWatchOnEC2Instance \\
14 & 33 & AWSConfigRemediation-EnableAutoScalingGroupELBHealthCheck \\
15 & 88 & AWSConfigRemediation-RemoveUnrestrictedSourceIngressRules \\
16 & 45 & AWS-CreateS3PolicyToExpireMultipartUploads \\
17 & 28 & AWS-EnableCloudTrailCloudWatchLogs \\
18 & 50 & AWS-DisablePublicAccessForSecurityGroup \\
19 & 44 & AWSConfigRemediation-SetIAMPasswordPolicy \\
20 & 30 & AWSConfigRemediation-ConfigureS3PublicAccessBlock \\
21 & 78 & AWSFleetManager-CreateUser \\
22 & 74 & AWS-ConfigureCloudTrailLogging \\
23 & 43 & AWSConfigRemediation-EnableAccountAccessAnalyzer \\
24 & 90 & AWS-ChangeDDBRWCapacityMode \\
25 & 62 & AWSSupport-AnalyzeEBSResourceUsage \\
26 & 82 & AWS-SetupInventory \\
27 & 54 & AWS-CreateDSManagementInstance \\
28 & 13 & AWS-SetRequiredTags \\
29 & 64 & AWS-ModifyDynamoDBProvisionedCapacity \\
30 & 22 & AWSConfigRemediation-DetachIAMPolicy \\
31 & 77 & AWSSupport-SetupConfig \\
32 & 17 & AWS-EnableCWAlarm \\
33 & 20 & AWS-EnableS3BucketEncryption \\
34 & 52 & AWSConfigRemediation-EnforceEC2InstanceIMDSv2 \\
35 & 85 & AWS-CloseSecurityGroup \\
36 & 36 & AWSSupport-SetupIPMonitoringFromVPC \\
37 & 7 & AWSConfigRemediation-EnablePITRForDynamoDbTable \\
38 & 37 & AWSConfigRemediation-ConfigureS3BucketPublicAccessBlock \\
39 & 63 & AWSConfigRemediation-DisableSubnetAutoAssignPublicIP \\
40 & 80 & AWSConfigRemediation-EnableCWLoggingForSessionManager \\
41 & 11 & AWS-RemediateSSMAgentVPCEndpoints \\
42 & 47 & AWS-EnableS3BucketKeys \\
\bottomrule
\label{tab:per_flask_csv_runbooks}
\end{tabular}
}
\end{table}

% \newpage 

% Requires: \usepackage{booktabs,array}
\begin{table}[t]
\centering
\caption{AWS Runbooks applicable to \texttt{live-score}}
\vspace{-3mm}
{\scriptsize
\begin{tabular}{ccp{9cm}}
\toprule
\textbf{No.} & \textbf{Idx} & \textbf{Scenario} \\
\midrule
1 & 6 & AWSConfigRemediation-EnableSecurityHub \\
2 & 10 & AWS-RestrictIncomingTraffic \\
3 & 74 & AWS-ConfigureCloudTrailLogging \\
4 & 43 & AWSConfigRemediation-EnableAccountAccessAnalyzer \\
5 & 66 & AWSConfigRemediation-EncryptSNSTopic \\
6 & 20 & AWS-EnableS3BucketEncryption \\
7 & 90 & AWS-ChangeDDBRWCapacityMode \\
8 & 52 & AWSConfigRemediation-EnforceEC2InstanceIMDSv2 \\
9 & 26 & AWSSupport-ConfigureEC2Metadata \\
10 & 79 & AWS-DisableIncomingSSHOnPort22 \\
11 & 32 & AWS-DisableEventBridgeRule \\
12 & 2 & AWS-EnableVPCFlowLogs \\
13 & 35 & AWS-DisableS3BucketPublicReadWrite \\
14 & 88 & AWSConfigRemediation-RemoveUnrestrictedSourceIngressRules \\
15 & 53 & AmazonCloudWatch-MigrateCloudWatchAgent \\
16 & 87 & AWSConfigRemediation-EncryptLambdaEnvironmentVariablesWithCMK \\
17 & 72 & AWSConfigRemediation-RestrictBucketSSLRequestsOnly \\
18 & 75 & AWSSupport-EnableVPCFlowLogs \\
19 & 28 & AWS-EnableCloudTrailCloudWatchLogs \\
20 & 50 & AWS-DisablePublicAccessForSecurityGroup \\
21 & 13 & AWS-SetRequiredTags \\
22 & 41 & AWSConfigRemediation-DeleteUnusedSecurityGroup \\
23 & 12 & AWS-CreateManagedLinuxInstance \\
24 & 86 & AWS-ConfigureS3BucketLogging \\
25 & 65 & AWS-ConfigureS3BucketVersioning \\
26 & 19 & AWSConfigRemediation-CreateCloudTrailMultiRegionTrail \\
27 & 44 & AWSConfigRemediation-SetIAMPasswordPolicy \\
28 & 30 & AWSConfigRemediation-ConfigureS3PublicAccessBlock \\
\bottomrule
\label{tab:per_live-score_csv_runbooks}
\end{tabular}
}
\end{table}

% \newpage 

% Requires: \usepackage{booktabs,array}
\begin{table}[t]
\centering
\caption{AWS Runbooks applicable to \texttt{mega-mesh}}
\vspace{-3mm}
{\scriptsize
\begin{tabular}{ccp{9cm}}
\toprule
\textbf{No.} & \textbf{Idx} & \textbf{Scenario} \\
\midrule
1 & 84 & AWSConfigRemediation-EnableCloudTrailEncryptionWithKMS \\
2 & 13 & AWS-SetRequiredTags \\
3 & 59 & AWSConfigRemediation-EnableSystemsManagerSessionManagerAuditLogsToS3 \\
4 & 78 & AWSFleetManager-CreateUser \\
5 & 18 & AWS-ArchiveS3BucketToIntelligentTiering \\
6 & 40 & AWSSupport-RemediateLambdaS3Event \\
7 & 51 & AWS-AttachExcludeConditionToS3DenyPolicies \\
8 & 47 & AWS-EnableS3BucketKeys \\
9 & 62 & AWSSupport-AnalyzeEBSResourceUsage \\
10 & 24 & AWSConfigRemediation-UpdateXRayKMSKey \\
11 & 65 & AWS-ConfigureS3BucketVersioning \\
12 & 72 & AWSConfigRemediation-RestrictBucketSSLRequestsOnly \\
13 & 37 & AWSConfigRemediation-ConfigureS3BucketPublicAccessBlock \\
14 & 46 & AWS-RemediateSSMAgentVPCAttributes \\
15 & 1 & AWS-EnableNeptuneDbBackupRetentionPeriod \\
16 & 20 & AWS-EnableS3BucketEncryption \\
17 & 56 & AWS-JoinDirectoryServiceDomain \\
18 & 28 & AWS-EnableCloudTrailCloudWatchLogs \\
\bottomrule
\label{tab:per_mega_mesh_csv_runbooks}
\end{tabular}
}
\end{table}

\clearpage
\pagebreak
\begin{lstlisting}[
    breaklines=true,
    frame=single,
    basicstyle=\small\ttfamily,
    caption={Prompt for assessing whether a runbook scenario is applicable to an IaC project.},
    label={lst:scenario-prompt}
]
You are an expert in Terraform and AWS infrastructure management. 
You need to determine if a Terraform scenario can be meaningfully applied to mutate the current AWS infrastructure.

Analyze both inputs carefully:
- <STATE>: Outputs from the command `terraform state list` describing the resources under the current AWS infrastructure
- <SCENARIO>: A Terraform scenario description that explains how to implement a specific configuration change or infrastructure mutation

Each scenario aims to mutate the terraform infrastructure in a meaningful way by:
- Adding new resources or configurations
- Modifying existing resource properties
- Removing or deleting resources
- Implementing security improvements
- Enhancing monitoring or logging capabilities
- Optimizing resource configurations

Determine if the scenario can meaningfully mutate this infrastructure by checking:
1. If the resources mentioned in the scenario exist in the current Terraform state or can be logically added/removed
2. If the scenario would result in a small change to the infrastructure, typically involves only few resources 
3. If the scenario's requirements are compatible with the existing infrastructure architecture
4. If the mutation can be safely implemented without creating security risks or breaking existing functionality
5. If the scenario adds value by improving security, performance, monitoring, operational capabilities, or cost optimization through resource removal

A scenario is applicable if it can meaningfully transform or enhance the existing infrastructure state through addition, modification, or removal of resources.

Output Format: 
<REASON>A concise explanation of why the scenario is or is not meaningfully applicable to mutate this infrastructure, focusing on the value and feasibility of the proposed changes</REASON>
<LABEL>Yes</LABEL> if the scenario can meaningfully mutate the infrastructure, <LABEL>No</LABEL> if not applicable
\end{lstlisting}

% \newpage 
\clearpage
\pagebreak

\subsection{Manually Crafted Mutations} 
\label{sec:appendix-custom-mutation}

For each project, we hand-crafted a small set of mutations to complement the automatically generated ones. The project-specific cases are summarized for \texttt{lab12} in Table~\ref{tab:custom-lab12}, \texttt{flask} in Table~\ref{tab:custom-flask}, \texttt{ssm3} in Table~\ref{tab:custom-ssm3}, and \texttt{live-score} in Table~\ref{tab:custom-live-score}. 
Each mutation was tested to ensure it was both deployable and revertible. 
The \texttt{mega-mesh} project was excluded from large-scale mutation testing due to high deployment costs, and only a small number of mutations were evaluated.

\clearpage
\pagebreak
\begin{table}[htbp]
\caption{Manually Crafted Mutations for the \texttt{lab12} project.}
\vspace{-3mm}
\centering
{\scriptsize
\begin{tabular}{cp{4.5cm}p{8cm}}
\toprule
\textbf{No.} & \textbf{Manually Crafted Mutations for \texttt{lab12}} & \textbf{Description} \\
\midrule
1 & \_delete\_route\_table\_associations & The mutation removes route table associations between subnets and route tables in an egress VPC network configuration, eliminating the connections that direct traffic flow between network components. \\ \addlinespace
2 & \_delete\_huge\_networks & The mutation removes route tables, route associations, and Transit Gateway route table associations, eliminating the networking infrastructure that connects the VPCs. \\ \addlinespace
3 & \_update\_tgroute2\_recreate & Changes the destination CIDR block for a transit gateway route from 10.0.0.0/8 to 11.0.0.0/8 and removes lifecycle create\_before\_destroy settings from security groups. \\ \addlinespace
4 & \_update\_sg\_name\_recreate & This mutation updates the name of a security group from "VPC\_B\_SG" to "VPC\_B\_SG-updated" and cause resource to be recreated. \\ \addlinespace
5 & \_scale\_up\_vm\_vpcs & This diff modifies the infrastructure to scale up from single instances to multiple instances in each VPC by adding count variables, removing hard-coded IP addresses, and updating resource references to support multiple instances. \\ \addlinespace
6 & \_delete\_eip\_NGW\_2 & The mutation removes a second Network Address Translation (NAT) gateway along with its associated elastic IP, route table, and route table association, eliminating redundancy in the network architecture. \\ \addlinespace
7 & \_delete\_update\_vpc\_ec2\_getw & The mutation consolidates networking infrastructure by removing a NAT gateway, limiting security group access to a specific IP range, removing lifecycle rules, and deleting an EC2 instance connect endpoint. \\ \addlinespace
8 & \_delete\_Bastion\_Endpoint\_VPC\_B & The change removes the EC2 instance connect endpoint for VPC B along with its security group lifecycle configurations. \\ \addlinespace
9 & \_rediect\_vpca\_and\_b\_for\_ec2 & Swaps the network interfaces between two instances and removes lifecycle management blocks from security groups. \\ \addlinespace
\bottomrule
\label{tab:custom-lab12}
\end{tabular}
}
\end{table}

\begin{table}[htbp]
\caption{Manually Crafted Mutations for the \texttt{flask} project.}
\vspace{-3mm}
\centering
{\scriptsize
\begin{tabular}{cp{4.5cm}p{8cm}}
\toprule
\textbf{No.} & \textbf{Manually Crafted Mutations for \texttt{flask}} & \textbf{Description} \\
\midrule
1 & \_delete\_autoscaling\_alarms\_table & The mutation removes the DynamoDB table used for storing autoscaling alarms configuration while preserving the core application infrastructure including networking, IAM roles, ECS cluster, and CloudTrail logging capabilities. \\ \addlinespace
2 & \_delete\_music\_table & The mutation removes the DynamoDB music table resource while keeping the Flask application that references it, potentially causing application errors as the table no longer exists for storing and retrieving music data. \\ \addlinespace
3 & \_update\_role\_name\_recreations & The mutation involves relocating a Flask microservices application with its associated infrastructure from a nested directory structure to the root directory, updating the IAM role name from "ecs\_service\_role" to "ecs\_service\_role\_updated". \\ \addlinespace
4 & \_update\_scaling\_ecs\_count & The mutation updates the desired capacity for an ECS (Elastic Container Service) cluster from 1 to 5 instances, increasing the application's scalability to handle more traffic. \\ \addlinespace
5 & \_delete\_aws\_route\_table\_associations & The mutation removed AWS route table associations while setting up a complete Flask microservice infrastructure with CloudTrail logging, VPC networking, ECS clusters, load balancing, and DynamoDB tables. \\ \addlinespace
6 & \_delete\_polcy2 & This mutation creates a Flask microservice application deployed on AWS with CloudTrail logging capabilities, removing duplicate endpoint functions in the app.py file. \\ \addlinespace
7 & \_delete\_aws\_network\_acl\_rules & The mutation removes Network Access Control List (NACL) rules from a network configuration, effectively removing the firewall rules that control traffic to and from subnets in a VPC infrastructure. \\ \addlinespace
8 & \_delete\_scaling\_policys & This mutation removes auto-scaling policies from an AWS infrastructure while setting up a Flask microservice application with necessary infrastructure components like networking, IAM roles, DynamoDB database, and CloudTrail monitoring. \\ \addlinespace
9 & \_delete\_aws\_route\_pub & The mutation recreates an application infrastructure by removing a public route from a network configuration, which could improve security by limiting external access to resources. \\ \addlinespace
10 & \_delete\_aws\_network\_acl & The mutation moves a complete Flask microservice infrastructure from multiple AWS-specific folders to a single root directory, consolidating the implementation of a web application with DynamoDB backend, network infrastructure, and monitoring components. \\ \addlinespace
11 & \_delete\_aws\_route\_private\_1 & The mutation removes an AWS route for a private subnet (private\_1), relocating all infrastructure files from various AWS configuration patterns into a simplified Flask microservice environment. \\ \addlinespace
12 & \_update\_cloudwatch\_logs\_enabled & The mutation sets up a Flask microservices application with CloudTrail logging enabled and configures CloudWatch Logs integration for monitoring and auditing purposes. \\ \addlinespace
\bottomrule
\label{tab:custom-flask}
\end{tabular}
}
\end{table}

\begin{table}[htbp]
\caption{Manually Crafted Mutations for the \texttt{ssm3} project.}
\vspace{-3mm}
\centering
{\scriptsize
\begin{tabular}{cp{4.5cm}p{8cm}}
\toprule
\textbf{No.} & \textbf{Manually Crafted Mutations for \texttt{ssm3}} & \textbf{Description} \\
\midrule
1 & \_add\_ec2\_detailed\_monitoring & This mutation adds detailed CloudWatch monitoring to EC2 instances, improving operational visibility with 1-minute granularity metrics for better performance analysis and faster issue detection. \\ \addlinespace
2 & \_update\_s3\_policy\_readonly & This mutation updates an S3 bucket policy to remove write permissions, establishing read-only access for improved security testing. \\ \addlinespace
3 & \_delete\_kms\_key\_s3\_encryption & This mutation deletes a customer-managed KMS key and downgrades S3 bucket encryption from KMS to standard AES256 to test security configuration drift scenarios. \\ \addlinespace
4 & \_delete\_iam\_instance\_profile & This mutation removed the IAM instance profile from EC2 instances, resulting in the loss of Systems Manager management capabilities and automated maintenance functions. \\ \addlinespace
5 & \_delete\_iam\_mw\_policy & This mutation successfully removed an IAM policy and its role attachment, reducing permissions for maintenance window operations. \\ \addlinespace
6 & \_delete\_ssh\_security\_rule & The mutation deletes a security group rule that previously allowed SSH access from a specific IP address, blocking remote SSH access to EC2 instances. \\ \addlinespace
7 & \_update\_s3\_lifecycle\_retention & The mutation shortened the retention period for SSM output logs in an S3 bucket from 90 days to 30 days, reducing storage costs but potentially affecting compliance and debugging capabilities. \\ \addlinespace
8 & \_update\_s3\_bucket\_versioning & This mutation adds versioning capabilities to a storage bucket to enhance data protection and improve recovery options against accidental deletions or modifications. \\ \addlinespace
9 & \_update\_s3\_encryption\_aes256 & The report summarizes the simplification of S3 bucket encryption from customer-managed KMS to AWS-managed AES256 to reduce operational complexity while maintaining data security. \\ \addlinespace
10 & \_delete\_s3\_lifecycle\_config & This mutation deleted the automatic log cleanup configuration, which will cause SSM output logs to accumulate indefinitely instead of being deleted after 90 days. \\ \addlinespace
11 & \_update\_maintenance\_window\_schedule & The mutation report describes a change in system maintenance frequency from daily to weekly execution on Sundays. \\ \addlinespace
12 & \_update\_ec2\_instance\_type & This mutation updates the instance type of EC2 virtual machines from a smaller to a larger size, increasing the compute capacity of the infrastructure. \\ \addlinespace
13 & \_update\_alb\_listener\_port\_8080 & The mutation updated an application load balancer's listening port from 80 to 8080 for standardization purposes, affecting both the load balancer configuration and associated security group rules. \\ \addlinespace
14 & \_delete\_ssm\_graceful\_reboot & This mutation removed a Systems Manager document used for graceful server rebooting and updated related automation tasks to reference a different document instead. \\ \addlinespace
15 & \_update\_ec2\_instance\_count & This modification reduced the number of EC2 instances from 3 to 2 by changing the instance count parameter, resulting in the removal of one instance and its associated load balancer target group attachment. \\ \addlinespace
16 & \_update\_iam\_user\_tags & This mutation updates tags on an IAM user to test tag change detection and reconciliation processes in infrastructure management. \\ \addlinespace
17 & \_update\_alb\_health\_check & This mutation updated an application load balancer's health check configuration to use more conservative thresholds and longer intervals, which could result in slower detection of health status changes. \\ \addlinespace
18 & \_update\_security\_group\_cidr\_corporate & The update expanded network access permissions from a single IP address to a broader corporate network range (10.0.0.0/16) for both SSH and ALB ingress traffic, improving operational flexibility while maintaining security boundaries. \\ \addlinespace
\bottomrule
\label{tab:custom-ssm3}
\end{tabular}
}
\end{table}

\begin{table}[htbp]
\caption{Manually Crafted Mutations for the \texttt{live-score} project.}
\vspace{-3mm}
\centering
{\scriptsize
\begin{tabular}{cp{4.5cm}p{8cm}}
\toprule
\textbf{No.} & \textbf{Manually Crafted Mutations for \texttt{live-score}} & \textbf{Description} \\
\midrule
1 & \_delete\_ssh\_security\_group & This mutation removed a custom security group allowing unrestricted SSH access and replaced it with a more restrictive default security group to improve the system's security posture. \\ \addlinespace
2 & \_delete\_lambda\_event\_source\_mapping & This mutation removed the connection between a DynamoDB stream and Lambda function, disabling the automatic triggering of processor creation when database changes occur. \\ \addlinespace
3 & \_update\_sns\_topic & The mutation enhanced a notification topic by adding encryption and a descriptive display name to improve security and usability of the scorecard update system. \\ \addlinespace
4 & \_update\_cloudwatch\_schedule & This modification changes the execution schedule of the get-scorecard-urls Lambda function from a seasonal weekend-based pattern to a consistent daily weekday morning schedule that runs year-round. \\ \addlinespace
5 & \_update\_sqs\_queue & The configuration changes extended the message retention period and visibility timeout for a queue to improve reliability and processing time for web notification messages. \\ \addlinespace
6 & \_delete\_api\_authorizer\_lambda & This mutation removed authentication from the notifications API Gateway by eliminating the API authorizer Lambda function and its associated resources, making all notification endpoints publicly accessible without authentication. \\ \addlinespace
7 & \_delete\_update\_sanity\_module & The mutation removed the external Sanity CMS integration functionality from a live scores system while preserving all core scoring functionality. \\ \addlinespace
8 & \_update\_lambda\_runtime\_nodejs16x & This mutation downgraded the Node.js runtime version from 18.x to 16.x for all 16 Lambda functions in a live scores system while maintaining operational continuity. \\ \addlinespace
9 & \_update\_dynamodb\_connections\_encryption & Added server-side encryption to the live-score-connections database table to protect stored connection data through encryption at rest. \\ \addlinespace
10 & \_delete\_cloudwatch\_event\_rule & The mutation removed the scheduled automatic triggering mechanism for the get-scorecard-urls Lambda function while maintaining the function itself, eliminating automated cricket scorecard URL fetching during the season. \\ \addlinespace
11 & \_delete\_sqs\_scorecard\_html & This mutation removed an unused SQS queue resource and updated all dependent components to handle its absence gracefully through conditional logic. \\ \addlinespace
12 & \_delete\_api\_gateway\_authorizer & This modification removed authentication requirements from the API Gateway notifications endpoints, enabling unauthenticated access to test security configuration drift scenarios. \\ \addlinespace
13 & \_delete\_api\_route & This mutation removed a WebSocket connection route from the API gateway to simulate and test how the system would respond when a critical connection endpoint becomes unavailable. \\ \addlinespace
14 & \_update\_dynamodb\_subscriptions\_ttl\_gsi & This update adds time-based record expiration and a date-based lookup index to a database table for improved query performance, automatic data cleanup, and cost optimization. \\ \addlinespace
15 & \_delete\_dynamodb\_connections & The report summarizes the removal of a DynamoDB table used for WebSocket connection tracking, which eliminated the ability to track and manage connection states while preserving core system functionality. \\ \addlinespace
16 & \_update\_cloudwatch\_log\_retention & The report summarizes a successful update to CloudWatch log group retention policies across multiple Lambda functions, extending the retention period from 14 to 30 days to improve log management and compliance capabilities. \\ \addlinespace
17 & \_delete\_s3\_scorecards\_bucket & This mutation removed the S3 scorecards bucket and related configurations from a live scores project, eliminating scorecard storage functionality while maintaining system stability through conditional logic. \\ \addlinespace
18 & \_update\_security\_group & The mutation restricts SSH access from the public internet to private network ranges only, improving security by reducing the attack surface. \\ \addlinespace
19 & \_update\_sqs\_encryption & The mutation adds encryption at rest to a web notification queue system using AWS KMS, enhancing message security without disrupting service. \\ \addlinespace
\bottomrule
\label{tab:custom-live-score}
\end{tabular}
}
\end{table}

\end{document}